\documentclass[letterpaper]{article}

\usepackage{natbib,alifeconf}  
\usepackage{graphicx}
\usepackage{amsmath}
\usepackage{amsfonts}
\usepackage{comment}
\usepackage{tikz}
\let\svtikzpicture\tikzpicture
\def\tikzpicture{\noindent\svtikzpicture}
\usepackage{graphicx}
\usetikzlibrary{calc}
\usepackage{array}
\usepackage{hyperref}
\usepackage{supertabular}

\def\N{\mathbb{N}}

\makeatletter 
\def\mod@estimate@lineht{%
  \ST@lineht=\arraystretch \baslineskp 
  \ST@stretchht\ST@lineht\advance\ST@stretchht-\baslineskp 
  \ifdim\ST@stretchht<\z@\ST@stretchht\z@\fi 
  \ST@trace\tw@{Average line height: \the\ST@lineht}%
  \ST@trace\tw@{Stretched line height: \the\ST@stretchht}%
} 
\newenvironment{strictsupertabular} 
  {\let\estimate@lineht\mod@estimate@lineht\supertabular} 
  {\endsupertabular} 
\makeatother

\title{Classification of Complex Systems Based on Transients}
\author{Barbora Hudcova$^{1, 2}$ \and Tomas Mikolov$^{2}$\\
\mbox{}\\
$^1$Charles University, Prague\\
$^2$Czech Institute of Informatics, Robotics and Cybernetics, CTU, Prague} 

\begin{document}
\maketitle

\begin{abstract}
In order to develop systems capable of modeling artificial life, we need
to identify, which systems can produce complex behavior. We present
a novel classification method applicable to any class of deterministic discrete space
and time dynamical systems. The method distinguishes between different
asymptotic behaviors of a system’s average computation time before
entering a loop. When applied to elementary cellular automata, we obtain
classification results, which correlate very well with Wolfram's manual
classification. Further, we use it to classify 2D cellular automata to show
that our technique can easily be applied to more complex models of
computation. We believe this classification method can help to develop
systems, in which complex structures emerge.
\end{abstract}

\section{Introduction}
There are many approaches to searching for systems capable of open-ended evolution. One option is to carefully design a model and observe its dynamics. Iconic examples were designed by \cite{tierra}, \cite{avida}, or \citet{chromaria}. However, as we lack any formal definition of open-endedness or complexity, there is no formal method of proving the system is indeed "interesting". Conversely, lacking definitions of such key terms, it seems extremely difficult to design such systems systematically. 

Approaching the problem of searching for open-endedness bottom up, we define a classification of deterministic discrete dynamical systems based on their asymptotic computation time with increasing space size. This method gives a surprisingly clear classification of the toy class of elementary cellular automata, which seems to correspond well to Wolfram's established yet informal four types of cellular automata dynamics. Subsequently, we use the classification to discover two-dimensional automata with emergent behavior. This demonstrates that the transient classification can be used to navigate us towards regions of interesting systems. Even though we are far from giving sufficient conditions for complexity, we hope this method helps us understand, which formally defined properties correlate with it. 

\section{Introducing Cellular Automata}
Informally, a cellular automaton (CA) can be perceived as a $k$-dimensional grid consisting of identical finite state automata. They are all updated synchronously in discrete time steps based on an identical update function depending only on the states of automata in their local neighborhood. A formal definition can be found in \cite{Kari_survey}.

CAs were first studied as models of self replicating structures  (\cite{neuman}, \cite{LANGTONselfrep}, \cite{REGGIAselfrep}). Subsequently, they were examined as dynamical systems (\cite{hedlund}, \cite{vichniac}, \cite{gutowitz_local}, \cite{statistical_complexity}), or as models of computation (\cite{toffoli}, \cite{mitchell_overview}). Being so simple to simulate, yet capable of complex behavior and emergent phenomena (\cite{crutchfield}, \cite{emergence1}), CAs provide a convenient tool to examine the key, yet undefined notions of complexity and emergence.

\subsection{Basic Notions}
We study the simple class of \textit{elementary cellular automata} (ECAs), which are one-dimensional CAs with two states $\{0, 1 \}$ and neighborhood of size 3. We examine the case of a finite cyclic grid of size $n \in \N$ and denote each ECA by the tuple $(\{0, 1 \}^n, F)$ where $F: \{0, 1 \}^n \rightarrow \{0, 1 \}^n$ is the global update rule.

We identify each local rule $f$ determining an ECA with the \textit{Wolfram number} of the ECA defined as:
$$2^0 f(0, 0, 0) + 2^1 f(0, 0, 1) + 2^2 f(0, 1, 0) + \ldots + 2^7 f(1, 1, 1).$$ 
We will refer to each ECA as a "rule $k$" where $k$ is the corresponding Wolfram number of its underlying local rule.

Given an ECA $(\{0, 1 \}^n, F)$ we define the trajectory of a configuration $u \in \{0, 1 \}^n$  as the sequence 
$$(u, F(u), F^2(u), \ldots).$$ 
The space-time diagram of such simulation is obtained by plotting the configurations as a horizontal rows of black and white squares (corresponding to states 1 and 0) with vertical axis determining the time, which is progressing downwards.

As the set of all grid configurations $\{0, 1 \}^n$ is finite, every trajectory $(u_0, u_1 = F(u_0), u_2 = F(u_1), \ldots)$ eventually becomes periodic. We call the preperiod of this sequence the \textit{transient of initial configuration $u$} and denote its length by $t_u$. More formally, we define $t_u$ to be the smallest positive integer $i$, for which there exist $j \in \N$, $j > i$, such that $F^i(u) = F^j(u)$. The periodic part of the sequence is called an \textit{attractor}. The \textit{phase space} of an ECA $(\{0, 1 \}^n, F)$ is a graph with vertices $V = \{0, 1 \}^n$ and edges $E = \{(u, F(u)), u \in \{0, 1 \}^n\}$. Such a graph is composed of components each containing one attractor and multiple transient paths leading to the attractor. The phase space completely characterizes the dynamics of the system, it is however infeasible to describe it for large $n$.

We note that properties of CA phase spaces were examined among others by \cite{wuensche_global}. Precisely for this purpose, a software was designed by \cite{ddlab}.

\hyphenation{ap-proximated}

\section{Cellular Automata Classifications} 
Even though there are many interesting definitions of complexity (\cite{chaitin}, \cite{bennett}, \cite{mcshea}), none of them seems to be perfectly suitable for studying complex systems. A crucial result helping us understand the notion of complexity in the context of dynamical systems would be a suitable classification of CAs, which would navigate us toward a region of CAs with complex behavior. An ideal classification would be based on a rigorously defined and easily measurable property. 

In this section we describe three qualitatively different classifications of ECAs and subsequently, we will compare our results to them.

\hyphenation{ap-proximated}

\subsection{Wolfram's Classification}
\begin{figure}[h!]
    \centering
    \begin{tikzpicture}[thick, every node/.style={inner sep=0,outer sep=0}]
  \node at (3.4, 0) {
     \includegraphics[width=0.43\linewidth]{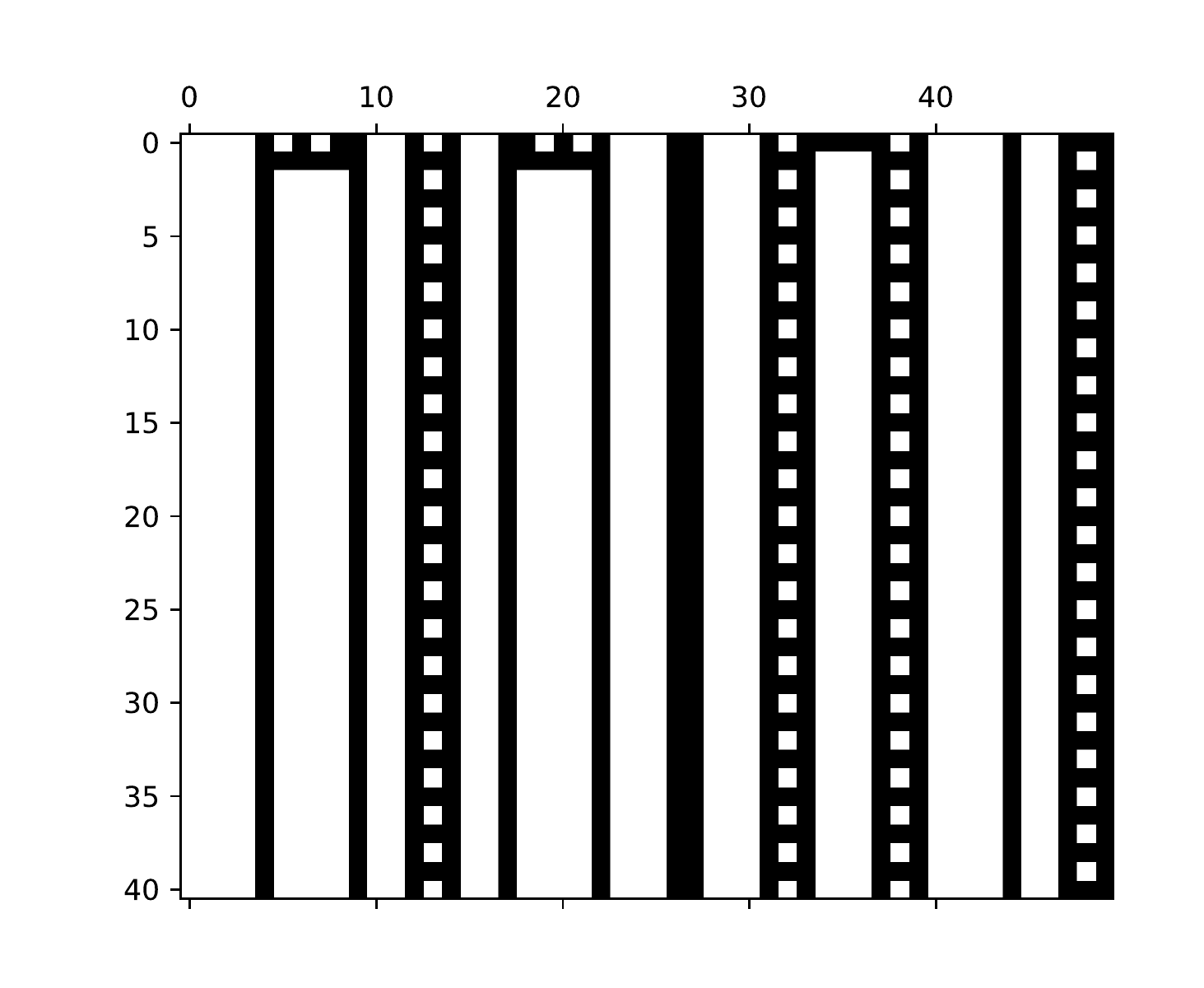}
  };
  \node at (0, 0) {
     \includegraphics[width=0.43\linewidth]{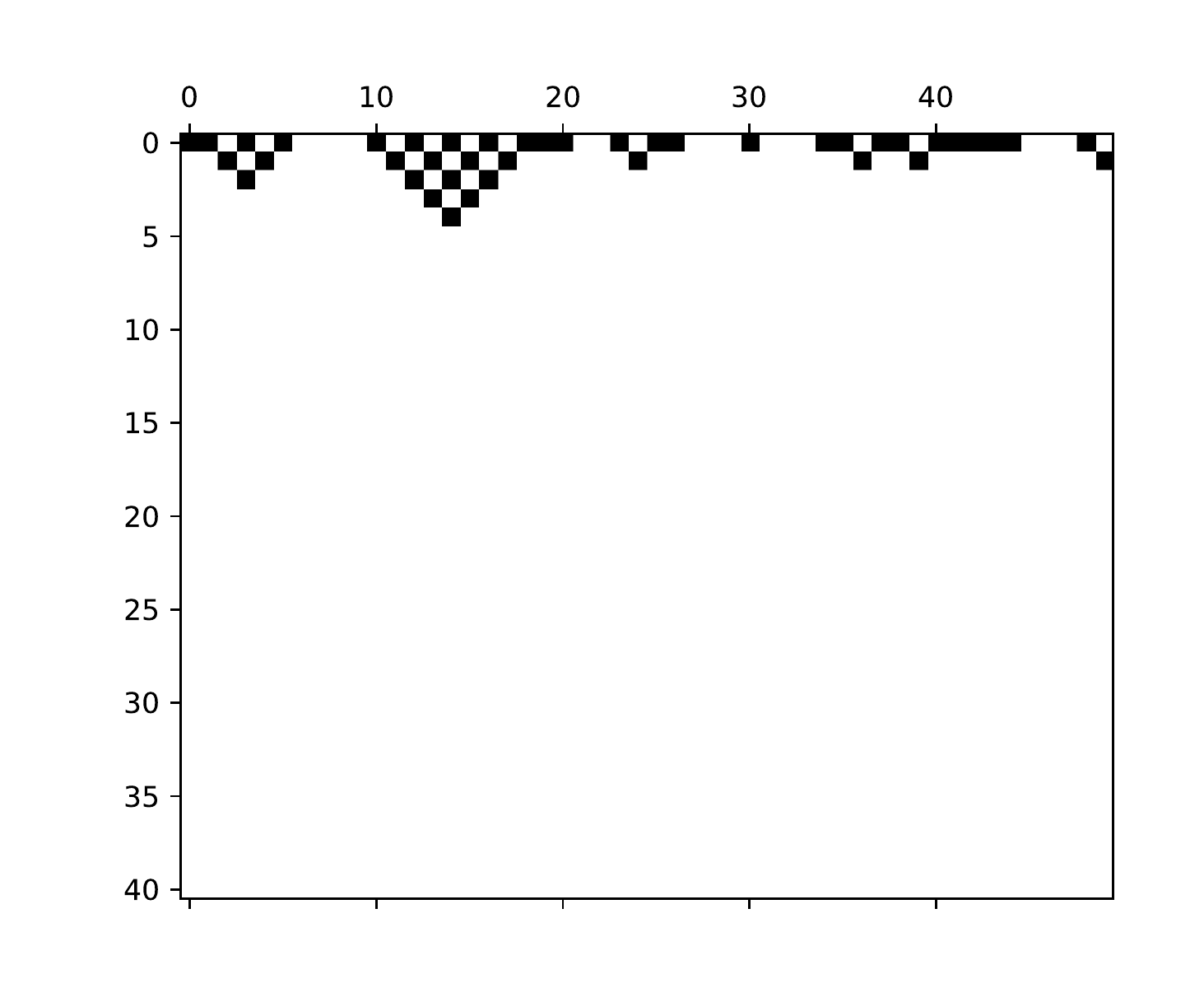}
  };
  \node at (3.4, -2.8) {
     \includegraphics[width=0.43\linewidth]{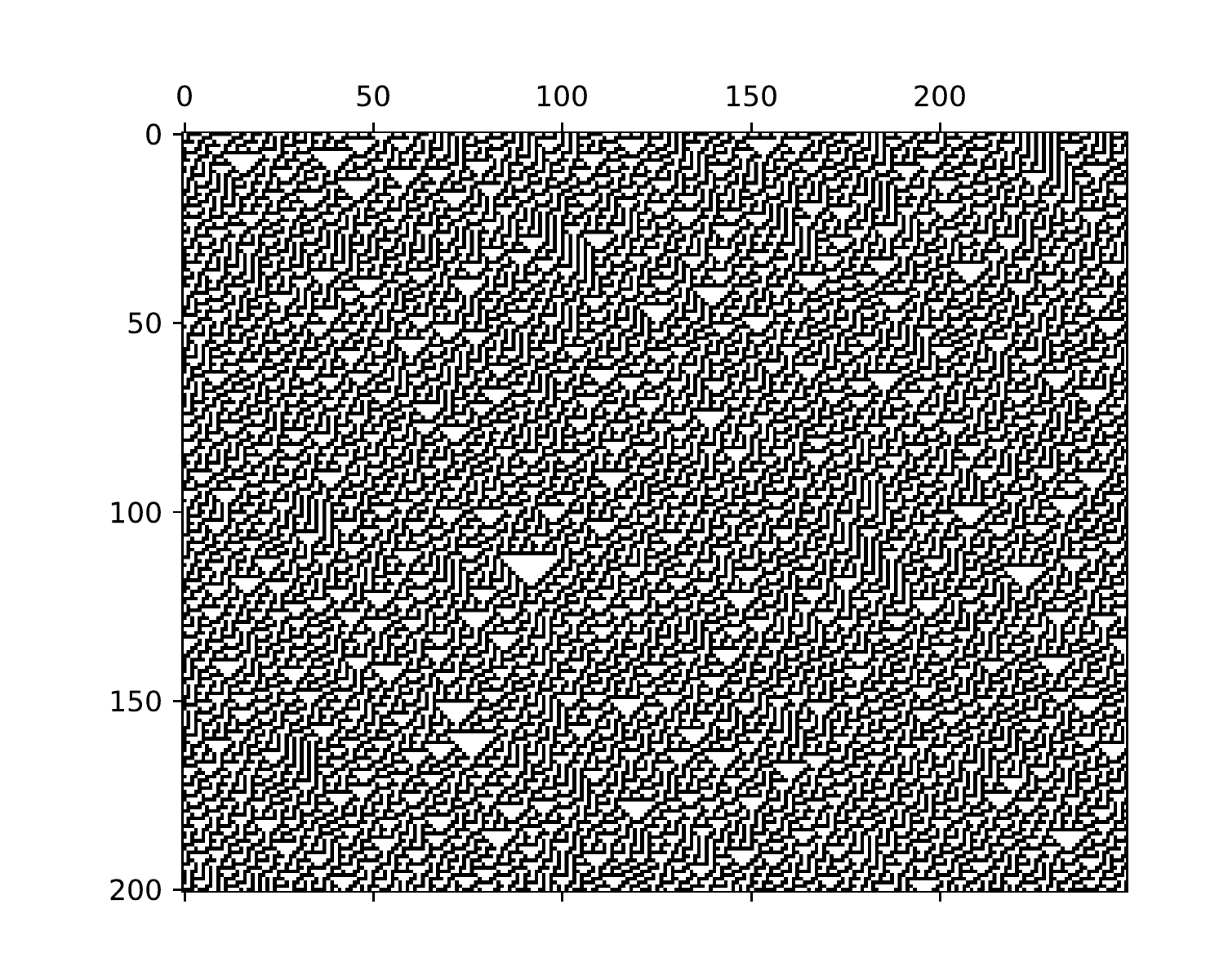}
  };
  \node at (0, -2.8) {
     \includegraphics[width=0.43\linewidth]{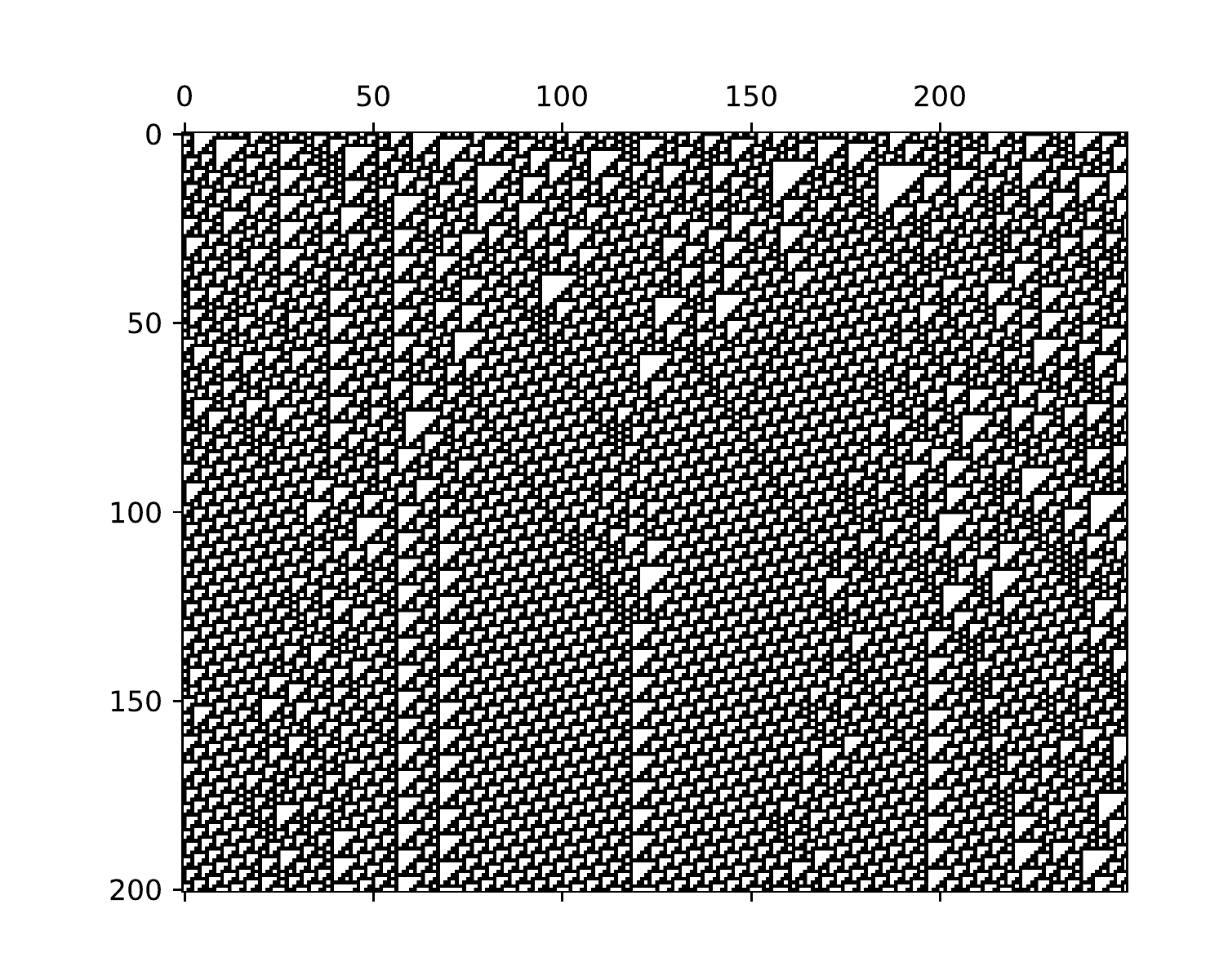}
  };
\end{tikzpicture}
    \caption{Space-time diagrams of rules from each Wolfram's class. Class 1 rule 32 is on top left, Class 2 rule 108 on top right. Both are simulated for 40 time steps on a grid of size 50. At the bottom row we have Class 4 rule 110 on the left and Class 3 rule 30 on the right. The two are simulated for 200 steps on a grid of size 250.}
    \label{wolf_classes}
\end{figure}

\begin{figure}[h!]
    \centering
    \begin{tikzpicture}[thick, every node/.style={inner sep=0,outer sep=0}]
  \node at (0, 0) {
     \includegraphics[width=0.5\linewidth]{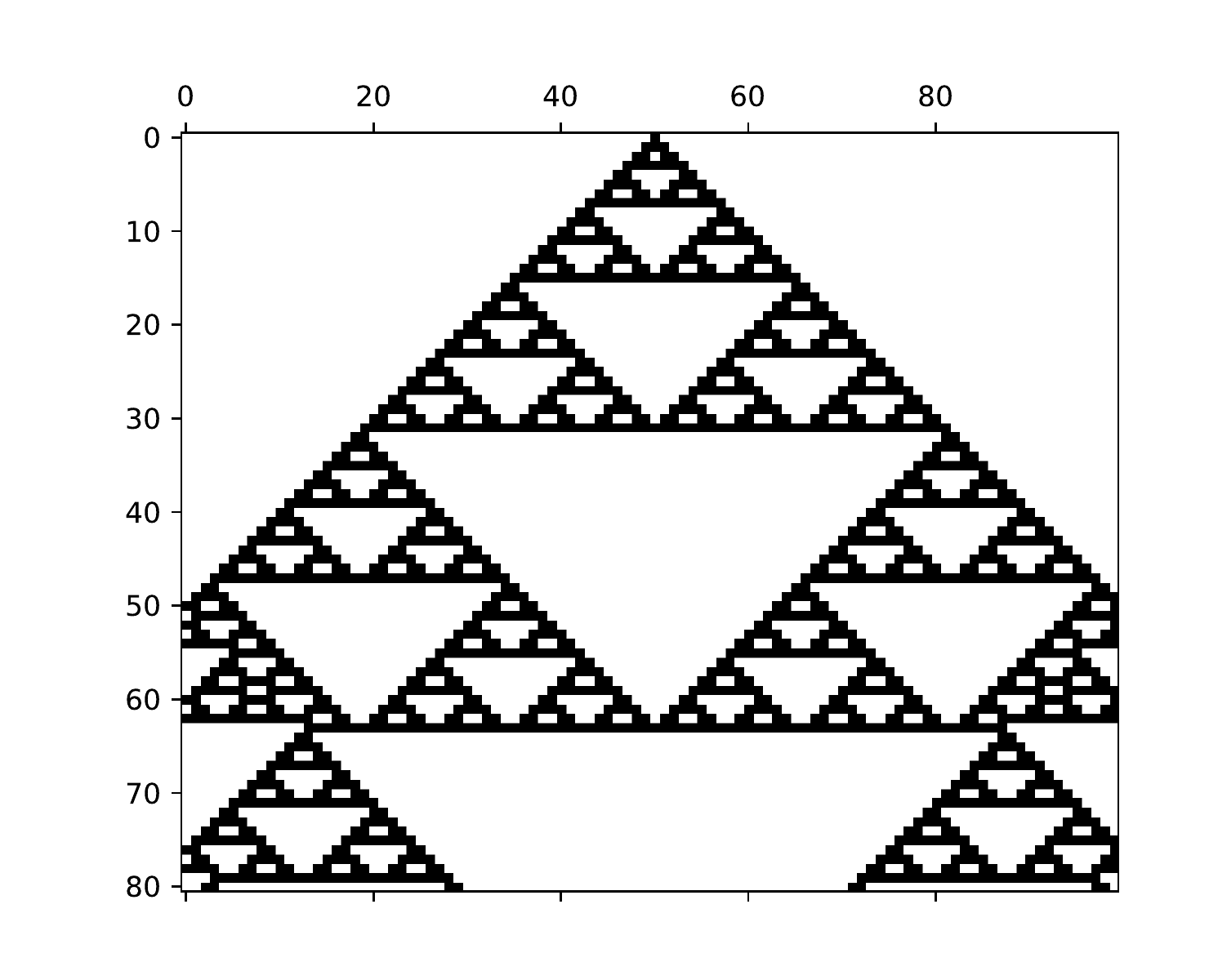}
  };
  \node at (4, 0) {
     \includegraphics[width=0.5\linewidth]{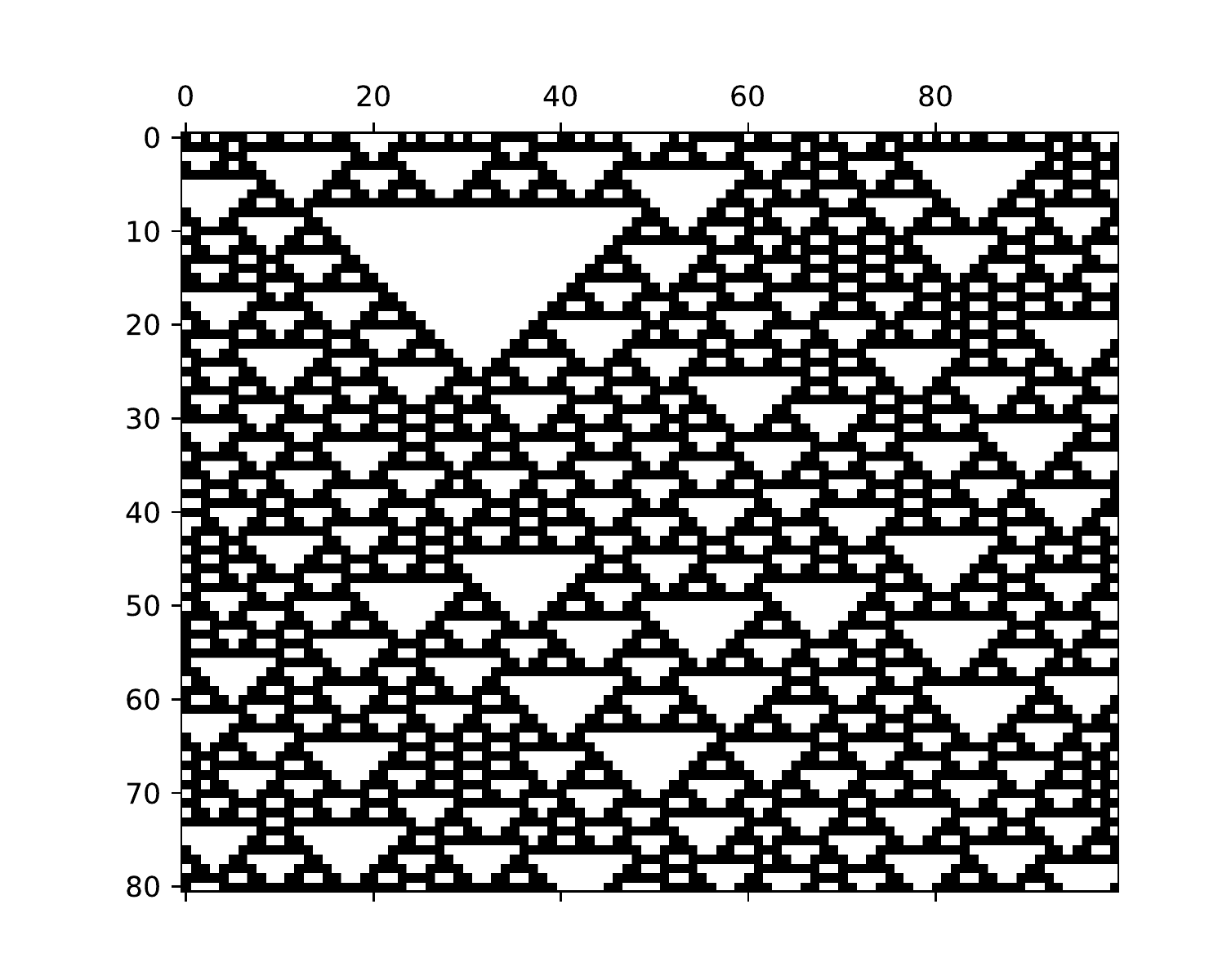}
  };
\end{tikzpicture}
\caption{On the left, rule 126 is simulated with an initial condition consisting of a single 1 bit padded with 0's. On the right, the same rule is simulated with a random initial configuration.}
\label{sensitiveCA}
\end{figure}

The most intuitive and simple approach to examining the dynamics of CAs is to observe their space-time diagrams. This method was particularly proclaimed by \cite{newscience}. Therein, he established an informal classification of CA dynamics based on such diagrams. He distinguishes the following classes, which are shown in Figure \ref{wolf_classes}.
\begin{align*}
\text{Class 1 } \ldots &\text{ quickly resolves to a homogenous state}\\
\text{Class 2 } \ldots &\text{ exhibits simple periodic behavior}\\
\text{Class 3 } \ldots &\text{ exhibits chaotic or random behavior}\\
\text{Class 4 } \ldots &\text{ produces localized structures that interact}\\
&\text{ with each other in complicated ways}
\end{align*}
The main issue is that we have no formal method of classifying CAs in this way. Moreover, the behavior of some CAs can vary with different initial configurations. An example being rule 126, which oscillates between Class 2 and Class 3 behavior, as shown in Figure \ref{sensitiveCA}. The transient classification we present in this paper deals with both these issues. 

\hyphenation{ap-proximated}

\subsection{Zenil's Classification}
\begin{figure}[h!]
    \centering
    \begin{tikzpicture}[thick, every node/.style={inner sep=0,outer sep=0}]
  \node at (1.5, 0) {
    \includegraphics[width=0.26\textwidth]{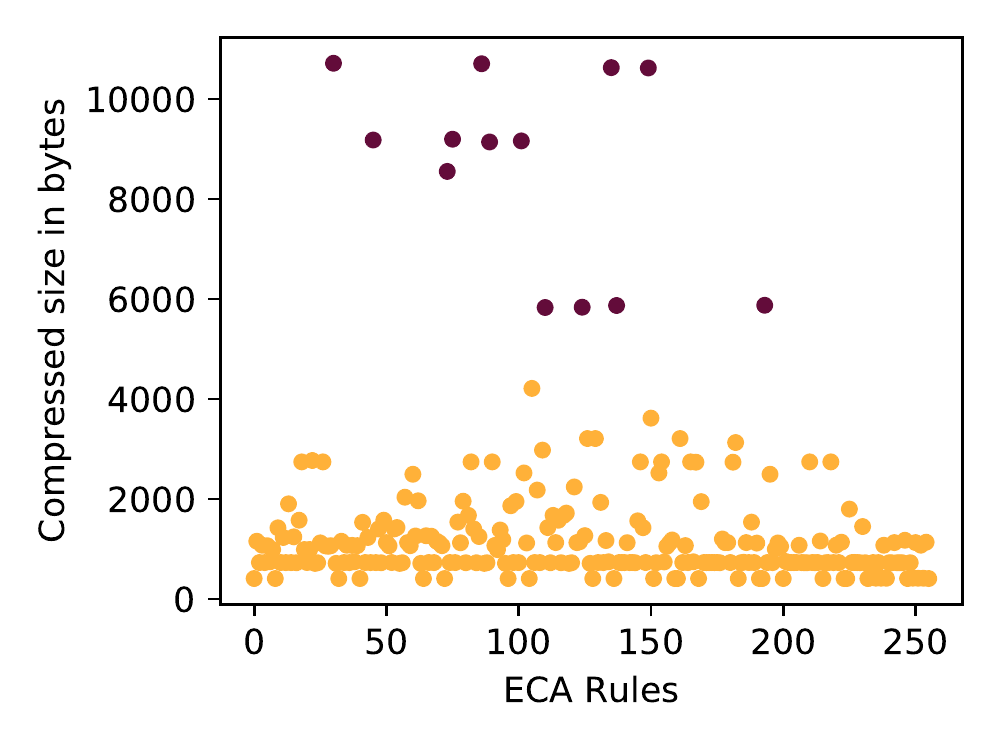}};
\end{tikzpicture}
\caption{Reproduction of Zenil's results (\cite{zenil_compression}). The purple cluster corresponds to the interesting Class 3 and 4 rules, the yellow cluster to the rest.}
\label{zenil_reproduction}
\end{figure}

\cite{zenil_compression} studied the compression size of the space-time diagrams of each ECA simulated for a fixed amount of steps. Using a clustering technique, he obtained two classes roughly distinguishing between Wolfram's simple classes 1 and 2 and complex classes 3 and 4. We show our reproduction of Zenil's results in Figure \ref{zenil_reproduction}.

\begin{figure}[h!]
    \centering
    \begin{tikzpicture}[thick, every node/.style={inner sep=0,outer sep=0}]

  \node at (0, 0) {
     \includegraphics[width=0.45\linewidth]{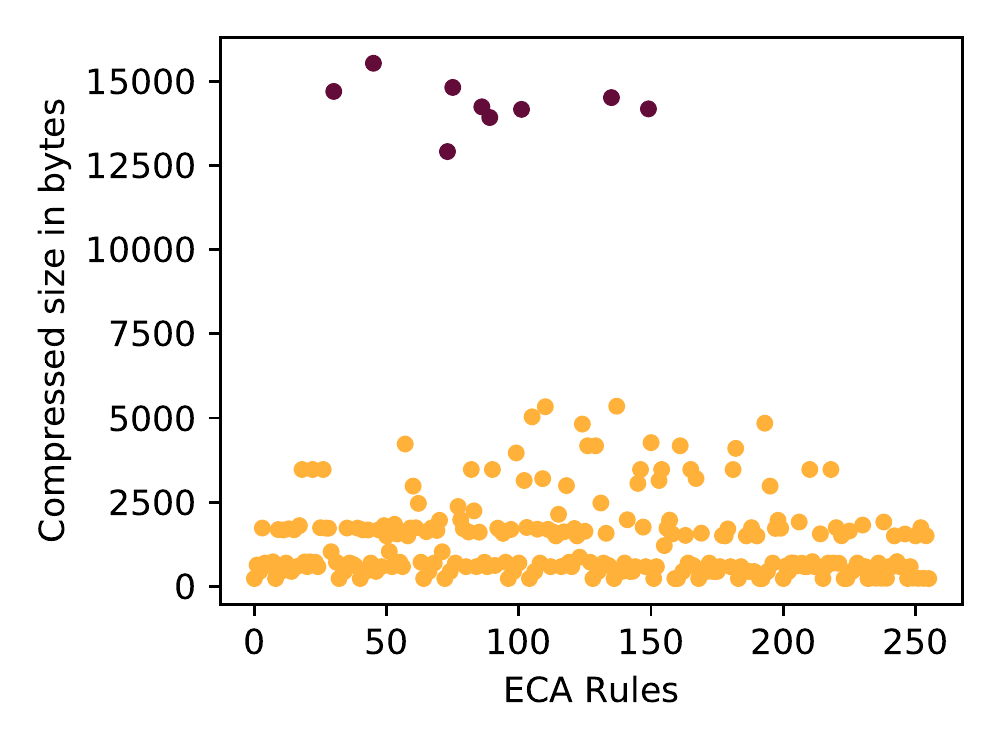}
  };
  \node at (3.9, 0) {
     \includegraphics[width=0.45\linewidth]{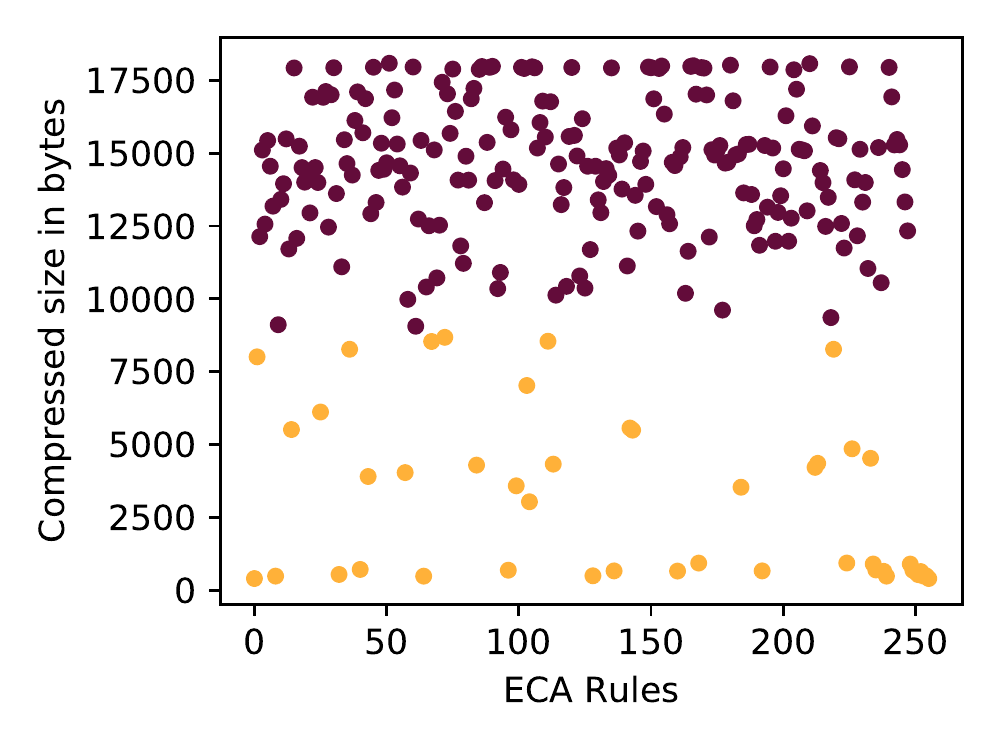}
 };
\end{tikzpicture}
\caption{Graphs representing the results of Zenil's method when different parameter values were used. They demonstrate how sensitive the results are. On the left, the ECAs were simulated for longer time, which caused complex rules 110, 124, 137, and 193 to no longer belong to the "interesting" purple cluster. On the right, the ECAs were simulated from a fixed, randomly chosen initial condition. In such case, we obtain entirely different clusters.}
\label{zenil_counter}
\end{figure}

His method nicely formalizes Wolfram's observations of the space-time diagrams. However, the results depend on the choice of initial conditions as well as the grid size, data representation, and the compression algorithm. We conducted multiple experiments presented in Figure \ref{zenil_counter}, which show that Zenil's results are very sensitive to the choice of such parameters. 

In vast CA spaces, where it is not feasible to examine every CA and mark it into one of Wolfram's classes by hand, it would not be clear how the parameter values should be chosen. Moreover, the data representation causes the extension of this method to more general dynamical systems to be problematic, as for example using gzip to compress space-time diagrams of a 2D cellular automaton is suboptimal.

\hyphenation{ap-proximated}

\subsection{Wuensche's Z-parameter}
In \cite{wuensche_global}, Wuensche chose an interesting approach by studying the ECA's behaviour when reversing the simulations and computing the preimages of each configuration. He introduces the Z-parameter, which represents the probability that a partial preimage can be uniquely prolonged by one symbol and suggests that Class 4 CAs typically occur at Z $\approx 0.75$. However, no clear classification is formed. The crucial advantage is that the Z-parameter depends only on the CA's local rule and can be computed effectively. It is, however, questionable whether studying the local rule only could describe overall dynamics of a system sufficiently well.

\subsection{Classification Based on Transients}

The classification we present is based on the asymptotic growth of the average transient length with increasing grid size. We will refer to it as the Transient Classification.
For a given CA and a grid size $n$, we randomly sample initial configurations $u \in \{0, 1\}^n$ and estimate the average transient length $\mu_n =\frac{1}{2^n} \sum_{u \in \{0, 1\}^n} t_u$. Using regression, we estimate the asymptotic growth of the sequence $(\mu_n)_{n=1}^\infty$. Below, we motivate the study of this property.

In non-classical models of compuation (\cite{natural_comp}), the process of traversing CA's transients can be perceived as the process of self-organization, in which information can be aggregated in an irreversible manner. The attractors are then viewed as memory storage units, from which the information about the output can be extracted. This is explored in \cite{kaneko}. Measuring the average transient growth then corresponds to the average computation time of the CA. CAs with bounded transient lengths can only perform trivial computation. On the other hand, CAs with exponential transient growth can be interpreted as inefficient computation models.

In the context of artificial evolution, we can view the local rule of a CA as the physical rule of the system whereas the initial configuration as the particular "setting of the universe", which is then subject to evolution. If we are interested in finding CAs capable of complex behavior automatically, it would be beneficial for us if such behavior occurred on average, rather than having to select the initial configurations carefully from some narrow region. The probability to find such special initial configurations would be extremely low as the overall number of configurations grows exponentially with increasing grid size. This motivates our study of the growth of average transient lengths rather than the maximum transient lengths.

We note that transients of CAs have been examined, as in \cite{wuensche_global} or \cite{gutowitz}. However, we are not aware of an attempt to compare the asymptotic growth of transients for different ECAs. 

\hyphenation{ap-proximated}

\section{Transient Classification of ECAs}

We consider all 256 ECAs up to equivalence classes obtained by changing the role of "left" and "right" neighbor, the role of 0 and 1 state, or both. 
It can be easily shown that automata in the same equivalence class have isomorphic phase spaces for any grid size. Thus, they perform the same computation. This yields 88 effectively different ECAs, each being a representative with the minimum Wolfram number from its corresponding equivalence class.
In this section, we present the classification of the 88 unique ECAs based on their asymptotic transient growth. First, we describe the details of the classification process.

\hyphenation{ap-proximated}

\subsection{Data Sampling and Regression Fits}
Suppose we have an ECA operating on a large grid of size $n$. In such case, computing the average transient length $\mu_n$ is infeasible. Therefore, we randomly sample initial configurations $u_1, u_2, \ldots, u_m$ and estimate $\mu_n$ by $\frac{1}{m} \sum_{i=1}^m t_{u_i}$. It remains to estimate the number of samples $m$ so that the error $|\frac{1}{m} \sum_{i=1}^m t_{u_i} - \mu_n|$ is reasonably small. 

More formally, we fix $n \in \N$ and let $(C_n, P_n)$ be a discrete probability space where $C_n = \{0, 1 \}^n$ is the set of all $n$-bit configurations and $P_n$ is a uniform distribution. Let $ X: C_n \rightarrow \N$ be a random variable, which sends each $u$ to its transient length $t_u$. This gives rise to a probability distribution of transient lengths on $\mathbb{N}$ with mean $E(X)$ and variance $var(X)$. It can be easily shown that $E(X) = \mu_n$. Our goal is to obtain a good estimate of $E(X)$ by the Monte Carlo method (\cite{mcbook}).

Let $(X_1, X_2, \ldots, X_m)$ be a random sample of iid random variables, $X_i \,{\buildrel d \over =}\, X$ for all $i$. Let $\mu_n^{(m)} = \frac{1}{m} \sum_{i=1}^m X_i$ be the sample mean and $\sigma_n^{(m)} = \sqrt{\frac{1}{m-1} \sum_{i=1}^m (X_i - \mu_n^{(m)} )^2}$ the sample standard deviation. As $var(X) < \infty$, we have by the Central limit theorem the convergence to a normal distribution, and the interval
$$\Big ( \mu_n^{(m)} - u_{1 - \frac{\alpha}{2}} \frac{\sigma_n^{(m)} }{\sqrt{m}}, \mu_n^{(m)} + u_{1 - \frac{\alpha}{2}} \frac{\sigma_n^{(m)} }{\sqrt{m}} \Big )$$
where $u_{\beta}$ is the $\beta$ quantile of the normalized normal
 distribution, covers $\mu_n$ for $m$ large with probability approximately $1-  \alpha$. We will take $\alpha = 0.05$. Hence, with probability approximately $95\%$ $$|\mu_n - \mu_n^{(m)}| < u_{0.975} \frac{\sigma_n^{(m)}}{\sqrt{m}}.$$
 
From the nature of our data, both the values $E(x) = \mu_n$ and $var(X)$ tend to grow with increasing grid size. Therefore, to employ a general method of estimating the number of samples, we normalize the error by the sample mean and consider $\frac{|\mu_n - \mu_n^{(m)}|}{\mu_n^{(m)}}$. 
 Therefore for $m$ sufficiently large such that 
 \begin{align} \label{sample_cond}
 u_{0.975}  \frac{\sigma_n^{(m)}}{\sqrt{m}  \mu_n^{(m)}} < \epsilon 
 \end{align}
 we have that $\mu_n^{(m)}$ differs from $\mu_n$ by at most $\epsilon \cdot 100\%$ with probability approximately $95\%$. 

In practice, we put $\epsilon = 0.1$ and produce the observations in batches of size 20 until condition (\ref{sample_cond}) is met.
For each ECA we obtained a dataset of the form $(\tilde{\mu}_{n})_{n = n_{min}}^{n_{max}}$ where $\tilde{\mu}_n$ is the estimate of the average transient length on the grid of size $n$. We typically put $n_{min} = 20$ and $n_{max} = 200$.

We examined different regression fits of the dataset to estimate the asymptotic growth of $\tilde{\mu}_n$. This included estimating the fit to constant, logarithmic, linear, polynomial, and exponential functions. We picked the best fit with respect to the $R^2$ score. Surprisingly, we found a very good fit with $R^2 > 90\%$ for most ECAs. 

\hyphenation{ap-proximated}

\subsection{Results}

We obtained a surprisingly clear classification of all the 88 unique ECAs with four major classes corresponding to the bounded, logarithmic, linear, and exponential growth of average transients. Below, we give a more detailed description of each class.

\paragraph*{Bounded Class:} 27/88 rules ($30.68\%$). The average transient lengths were bounded by a constant independent of the grid size. This suggests that the long term dynamics of such automata can be predicted efficiently.

\begin{figure}[h!] \label{bounded}
\begin{tikzpicture}[thick, every node/.style={inner sep=0,outer sep=0}]
  \node at (4, +0.08) {
     \includegraphics[width=0.44\linewidth]{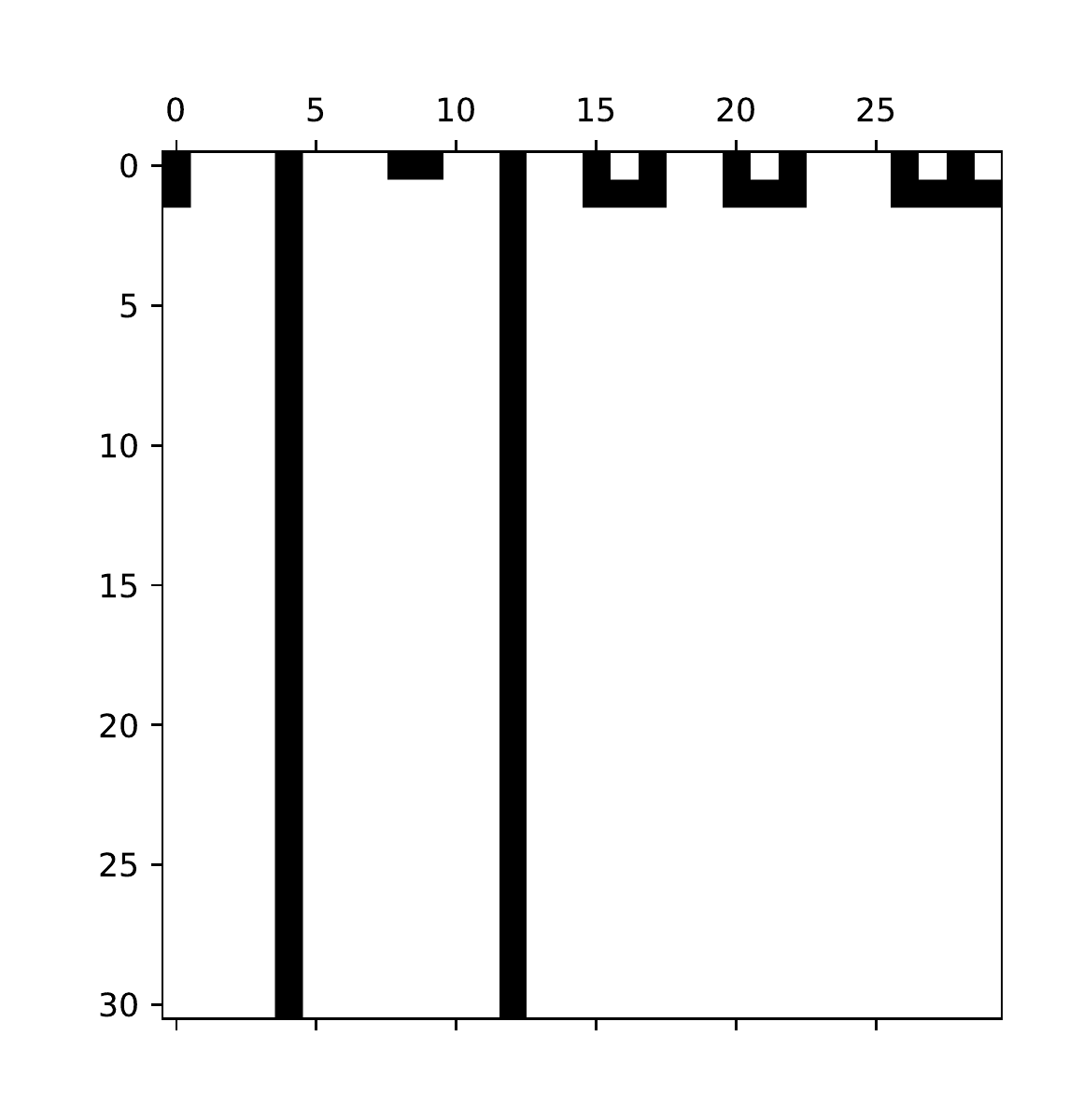}
  };
  \node at (-0.1, -0.1) {
     \includegraphics[width=0.6\linewidth]{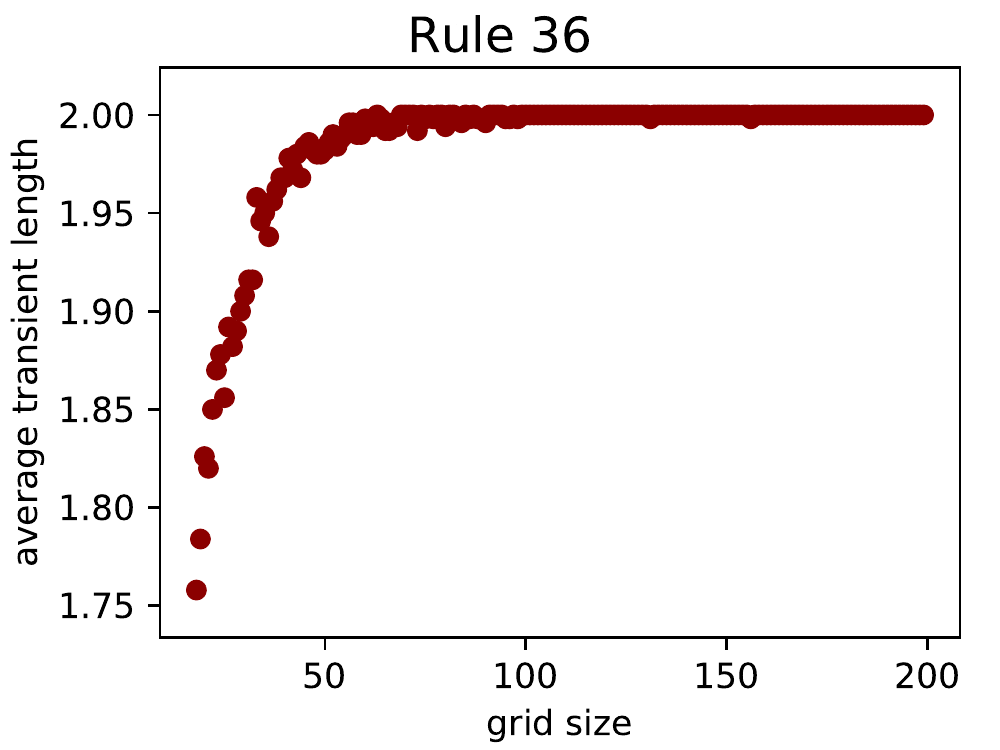}
  };
\end{tikzpicture}
\caption{Bounded Class rule 36. The average transient plot is on the left, the space-time diagram on the right.}
\end{figure}

\paragraph*{Log Class:}39/88 rules ($44.32\%$). The largest ECA class exhibits logarithmic average transient growth. The event of two cells at large distance "communicating" is improbable for this class. 

\begin{figure}[h!] \label{log}
    \centering
    \begin{tikzpicture}[thick, every node/.style={inner sep=0,outer sep=0}]
  \node at (4, +0.08) {
     \includegraphics[width=0.44\linewidth]{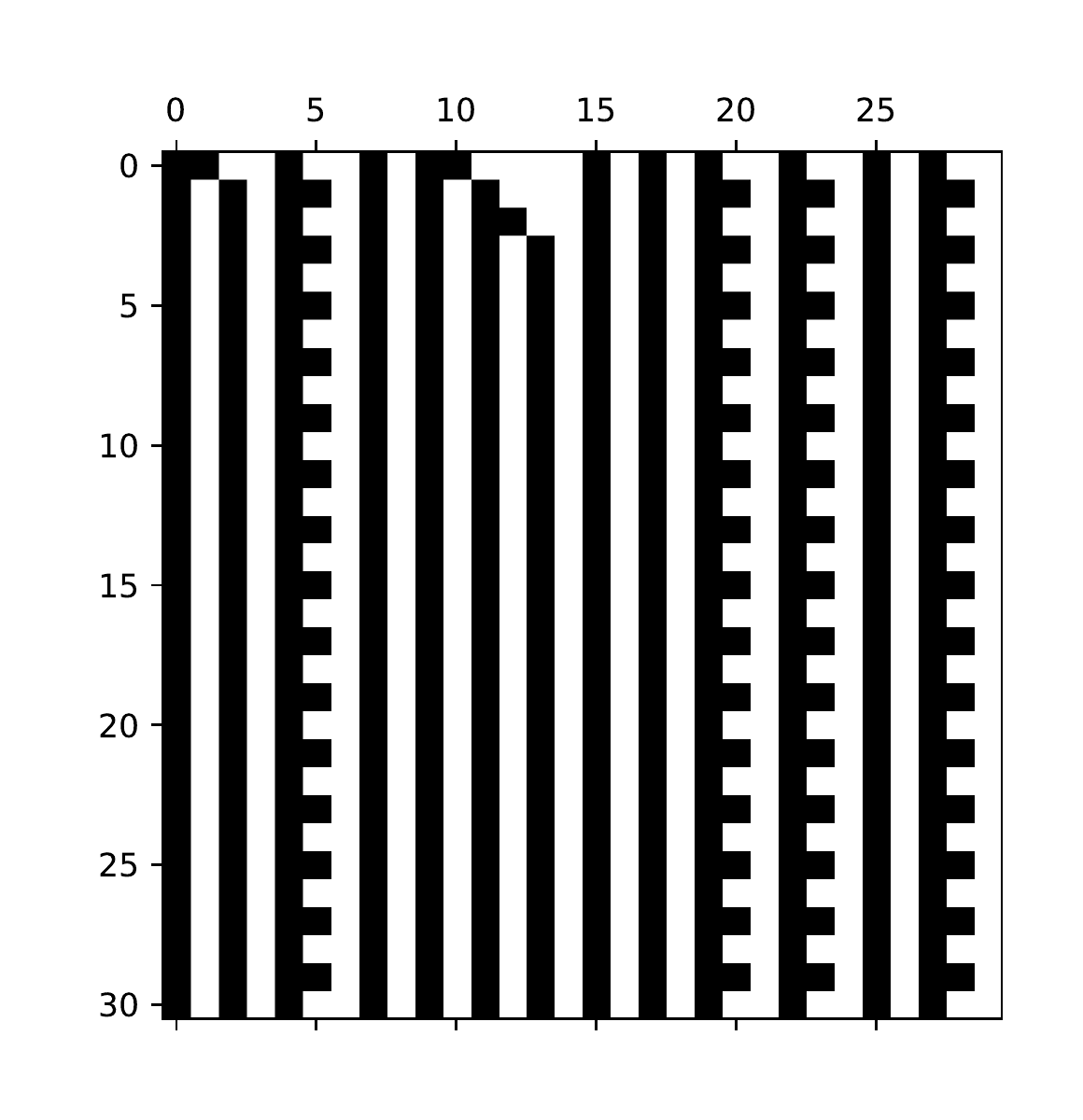}
  };
  \node at (-0.1, -0.1) {
     \includegraphics[width=0.6\linewidth]{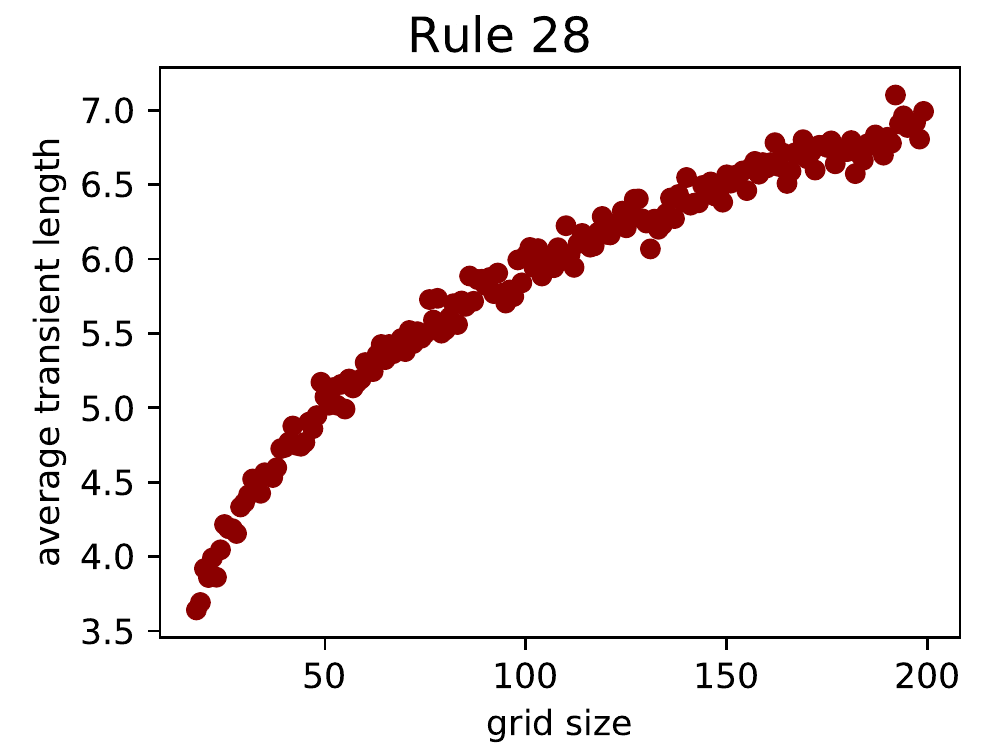}
  };
\end{tikzpicture}
\caption{Log Class rule 28. The average transient plot is on the left, the space-time diagram on the right.}
\end{figure}{}

\paragraph*{Lin Class:}8/88 rules ($9.09\%$). On average, information can be aggregated from cells at arbitrary distance. This class contains automata whose space-time diagrams resemble some sort of computation. This is supported by the fact that this class contains two rules known to have a nontrivial computational capacity: rule 184, which computes the majority of black and white cells, and rule 110, which is the only ECA proven to be Turing complete (\cite{cook}).

We note that rules in this class are not necessarily complex as the interesting behavior seems to correlate with the slope of the linear growth. Most of the Class Lin rules had only a very gradual incline. In fact, the only two rules with such slope greater than 1, rules 110 and 62, seem to be the ones with the most interesting space-time diagrams.

\begin{figure}[h!] \label{lin}
\begin{tikzpicture}[thick, every node/.style={inner sep=0,outer sep=0}]
  \node at (4, +0.08) {
     \includegraphics[width=0.44\linewidth]{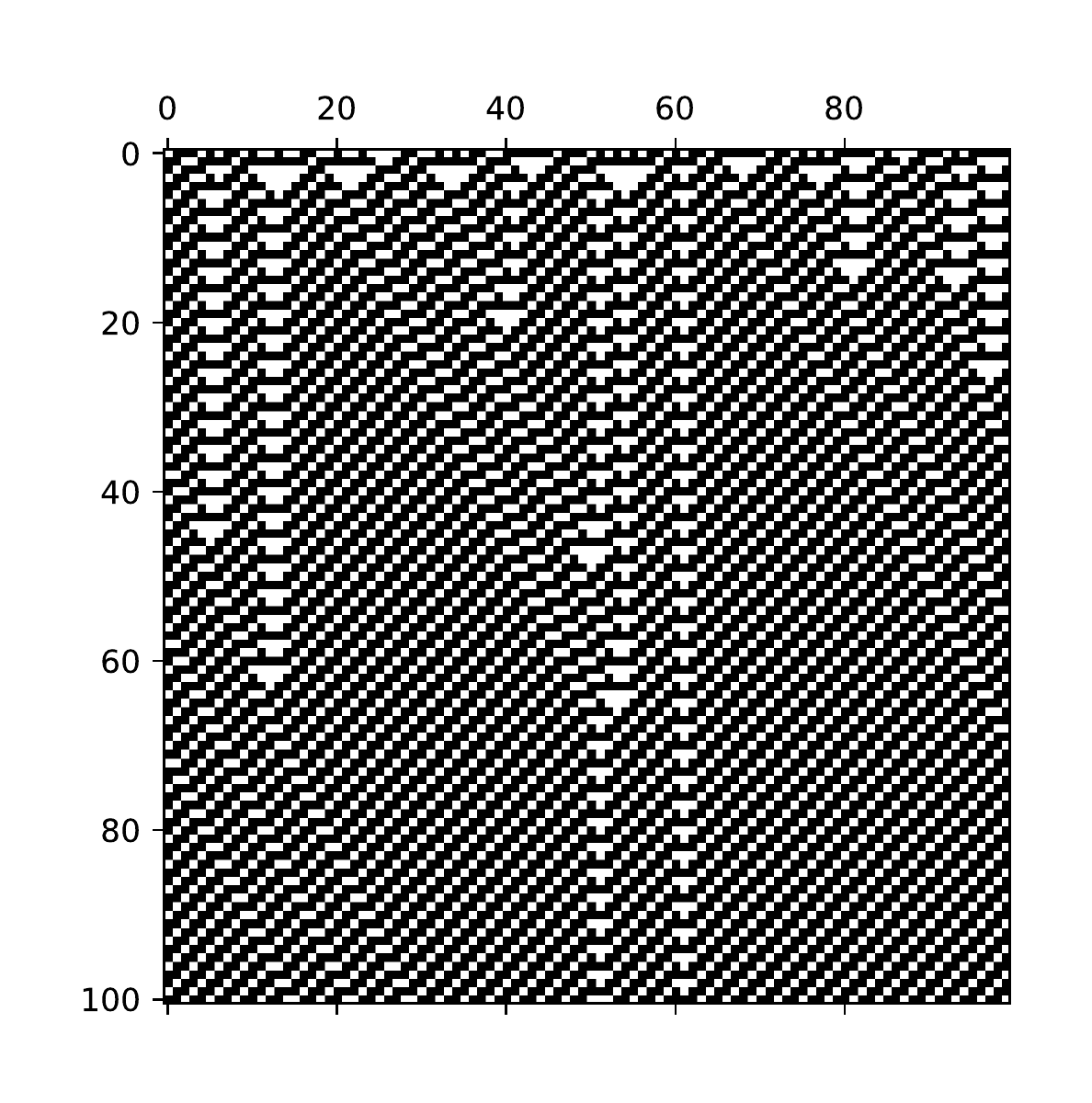}
  };
  \node at (-0.215, -0.1) {
     \includegraphics[width=0.6\linewidth]{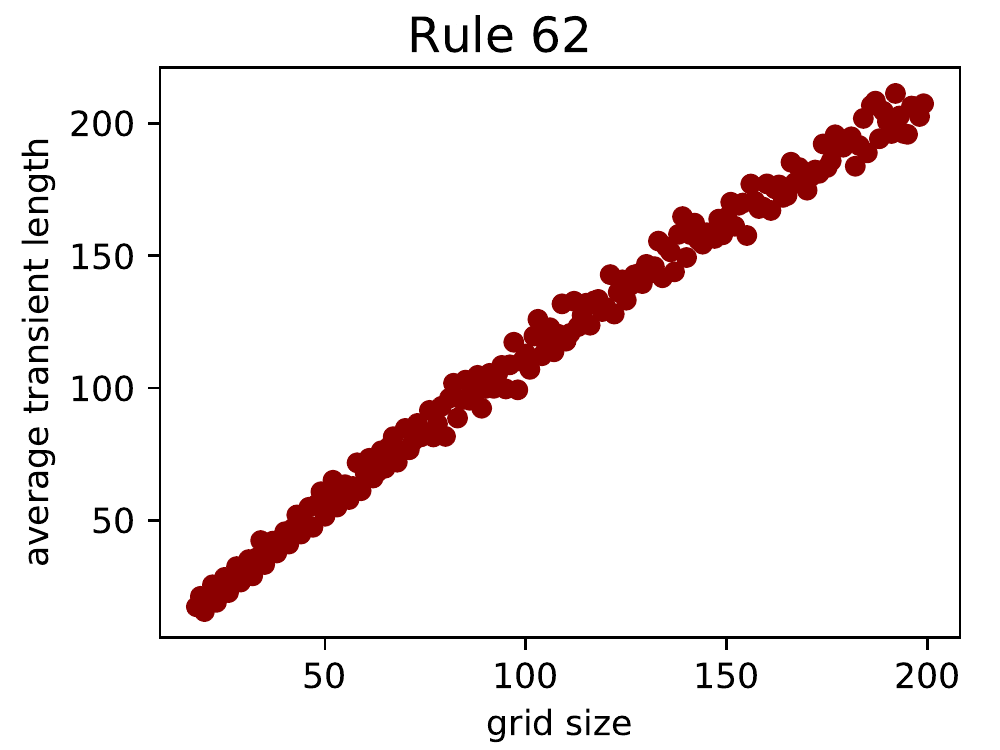}
  };
\end{tikzpicture}
\caption{Lin Class rule 62. The average transient plot is on the left, the space-time diagram on the right.}
\end{figure}

We are aware that average transients of rules in Lin Class might turn out to grow logarithmically or exponentially given enough data samples. This could explain why the behavior of ECAs in Lin Class depends on the slope of the transient growth. More formally, the class could be interpreted as consisting of rules, which might have a logarithmic or exponential growth, but this could not be decided given only a limited amount of data points. However, given such limited data, the best fit for such rules is to a linear function. 

\paragraph*{Exp Class:}6/88 rules ($6.82\%$). This class has a striking correspondence to automata with chaotic behavior. Visually, there seem to be no persistent patterns in the configurations. Not only the transients, but also the attractor lengths are significantly larger than for other rules. This class contains rules 45, 30 and 106 whose transients grow the fastest, as well as rules 54, 73, and 22.

\begin{figure}[h!] \label{exp}
\begin{tikzpicture}[thick, every node/.style={inner sep=0,outer sep=0}]
  \node at (4, +0.08) {
     \includegraphics[width=0.44\linewidth]{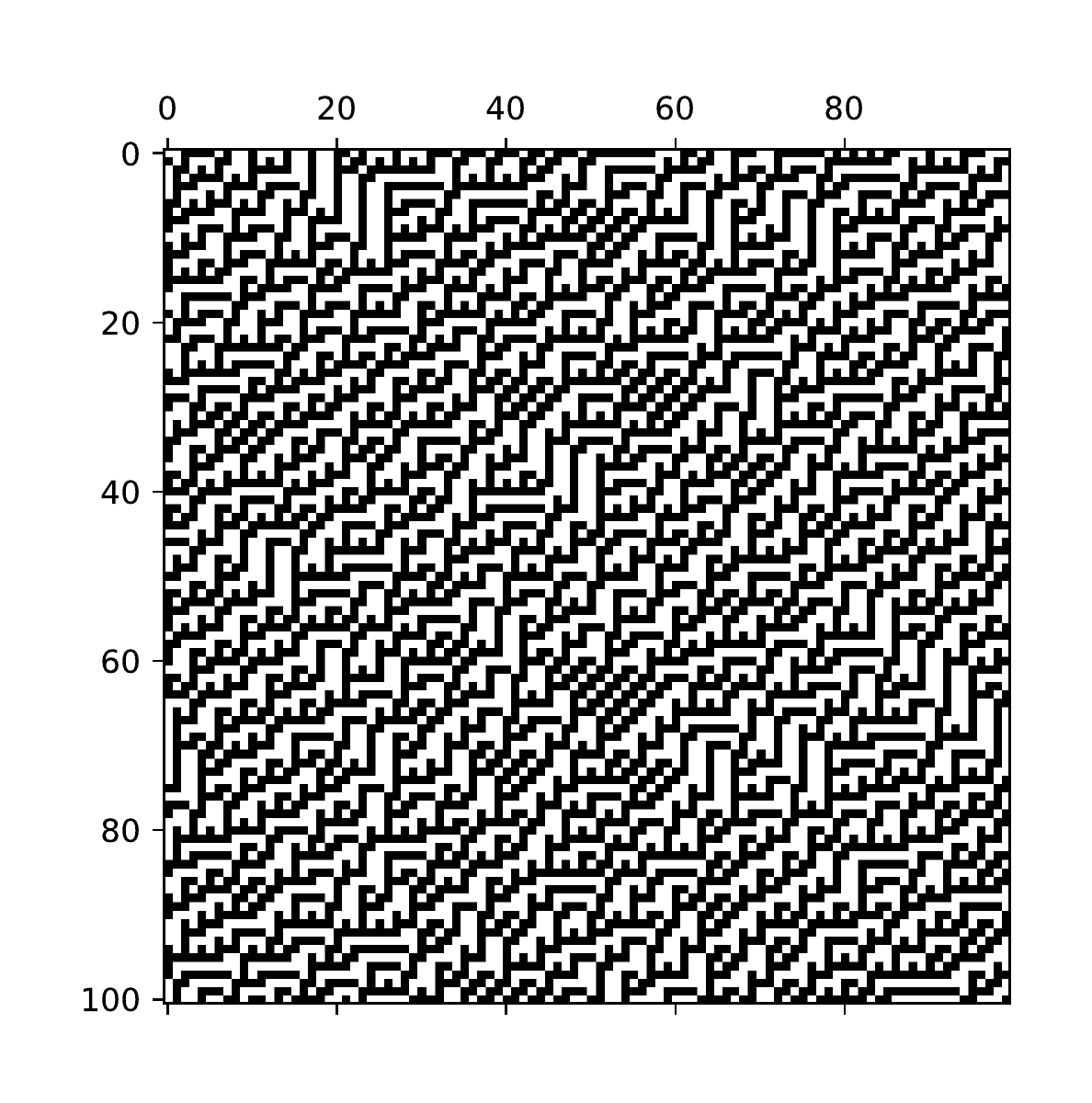}
  };
  \node at (-0.215, -0.1) {
     \includegraphics[width=0.6\linewidth]{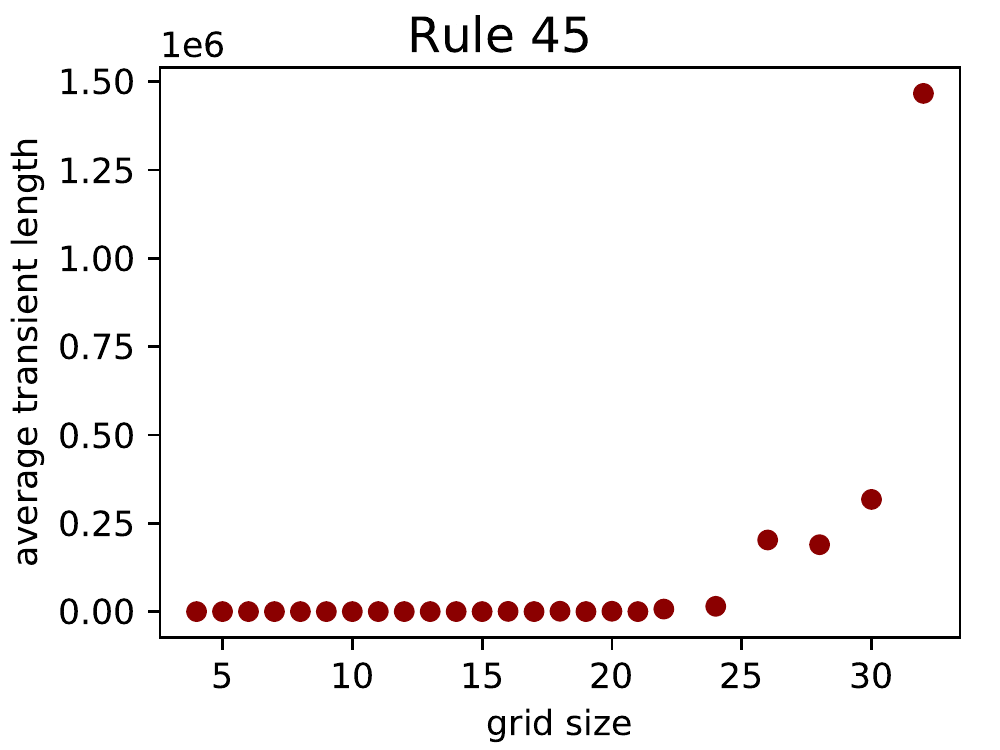}
  };
\end{tikzpicture}
\caption{Exp Class rule 45. The average transient plot is on the left, the space-time diagram on the right.}
\end{figure}

\paragraph*{Affine Class:}4/88 rules ($4.55\%$). This class contains rules 60, 90, 105, and 150 whose local rules are affine boolean functions. Such automata can be studied algebraically and predicted efficiently. It was shown in \cite{algCAs} that the transient lengths of rule 90 depend on the largest power of 2, which divides the grid size. Therefore, the measured data did not fit any of the functions above but formed a rather specific pattern.

\begin{figure}[h!] \label{affine}
\begin{tikzpicture}[thick, every node/.style={inner sep=0,outer sep=0}]
  \node at (4, +0.08) {
     \includegraphics[width=0.44\linewidth]{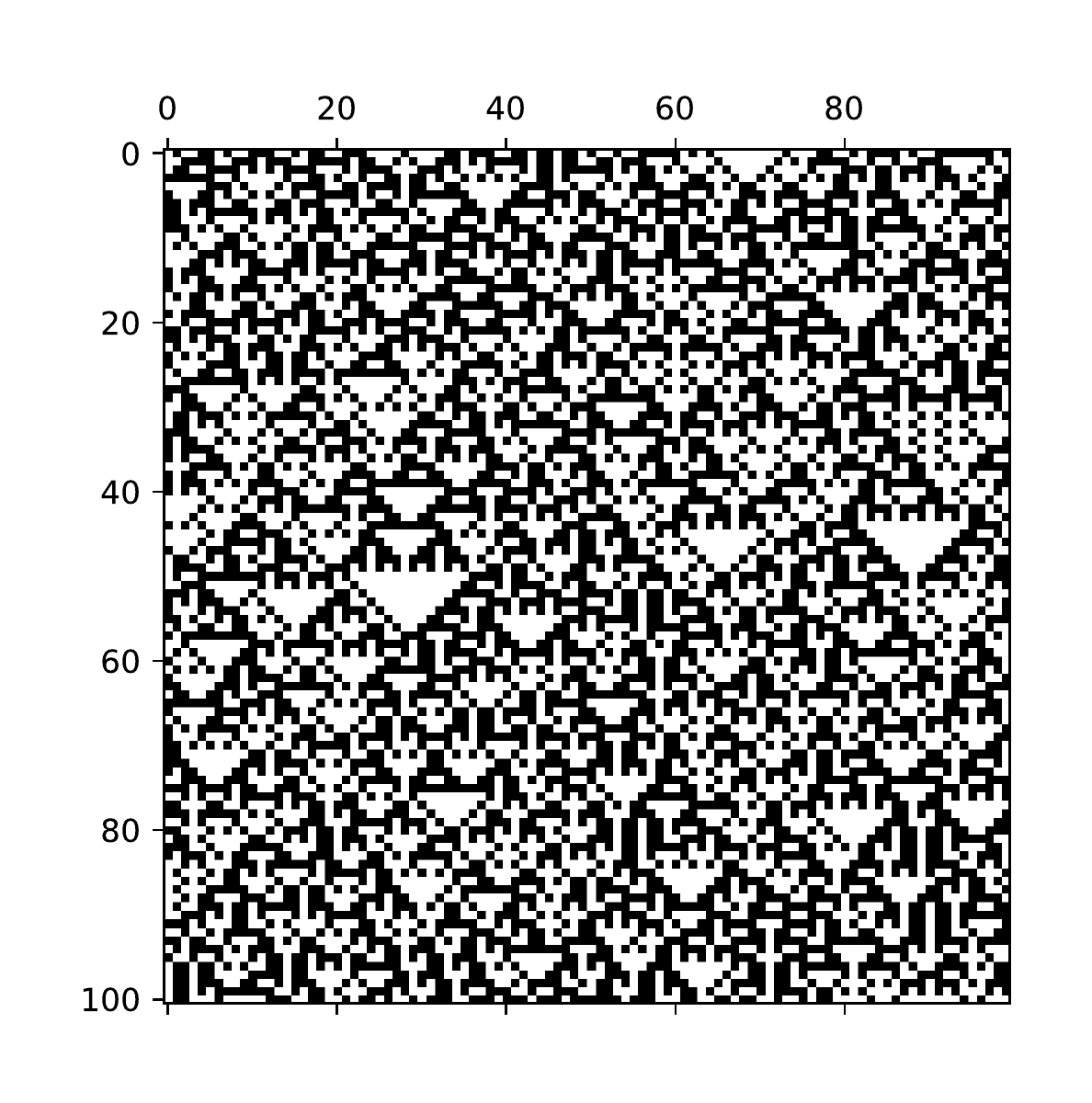}
  };
  \node at (-0.215, -0.1) {
     \includegraphics[width=0.6\linewidth]{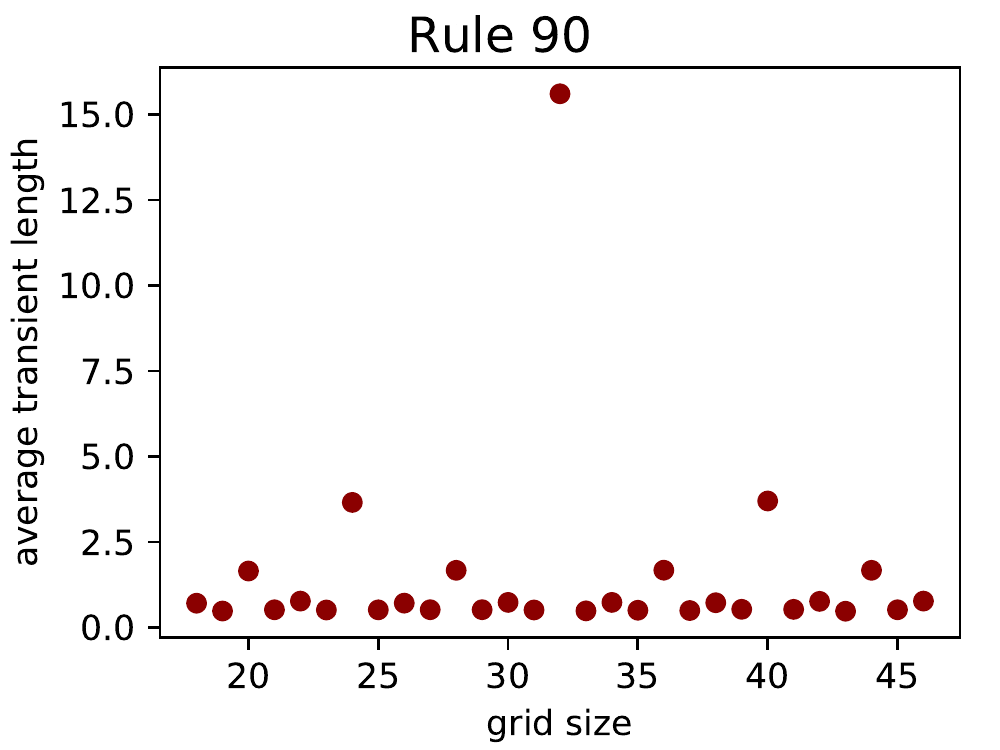}
  };
\end{tikzpicture}
\caption{Affine Class rule 90. The average transient plot is on the left, the space-time diagram on the right.}
\end{figure}

\paragraph*{Fractal Class:}4/88 rules  ($4.55\%$). This class contains rules 18, 122, 126, and 146, which are sensitive to initial conditions. Their evolution either produces a fractal structure resembling a Sierpinski triangle or a space-time diagram with no apparent structures. We could say such rules oscillate between easily predictable behavior and chaotic behavior. Their average transients and periods grow quite fast, which makes it difficult to gather data for larger grid sizes. 

\begin{figure}[h!] \label{fractal}
\begin{tikzpicture}[thick, every node/.style={inner sep=0,outer sep=0}]
  \node at (4, +0.08) {
     \includegraphics[width=0.44\linewidth]{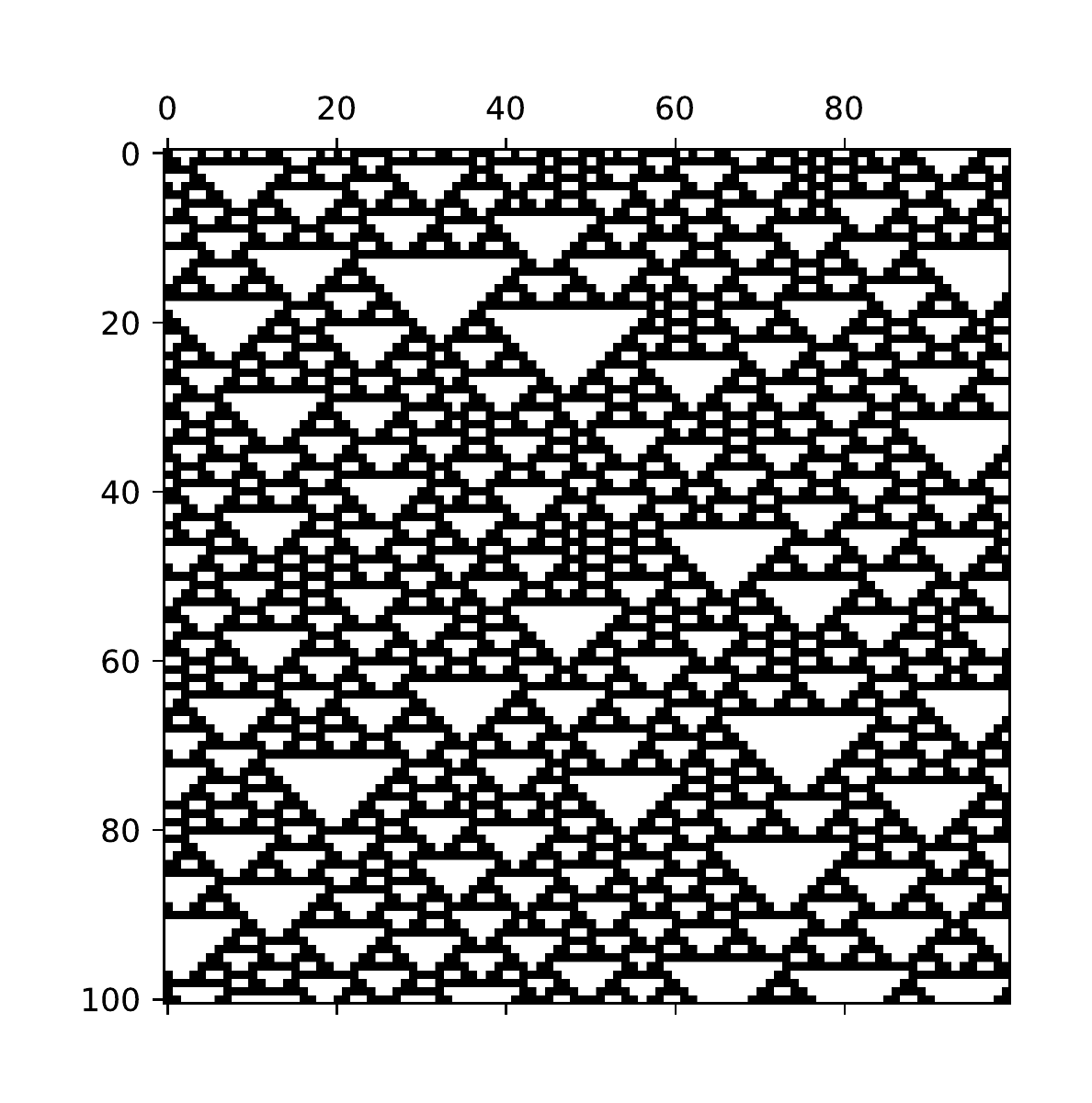}
  };
  \node at (-0.215, -0.1) {
     \includegraphics[width=0.6\linewidth]{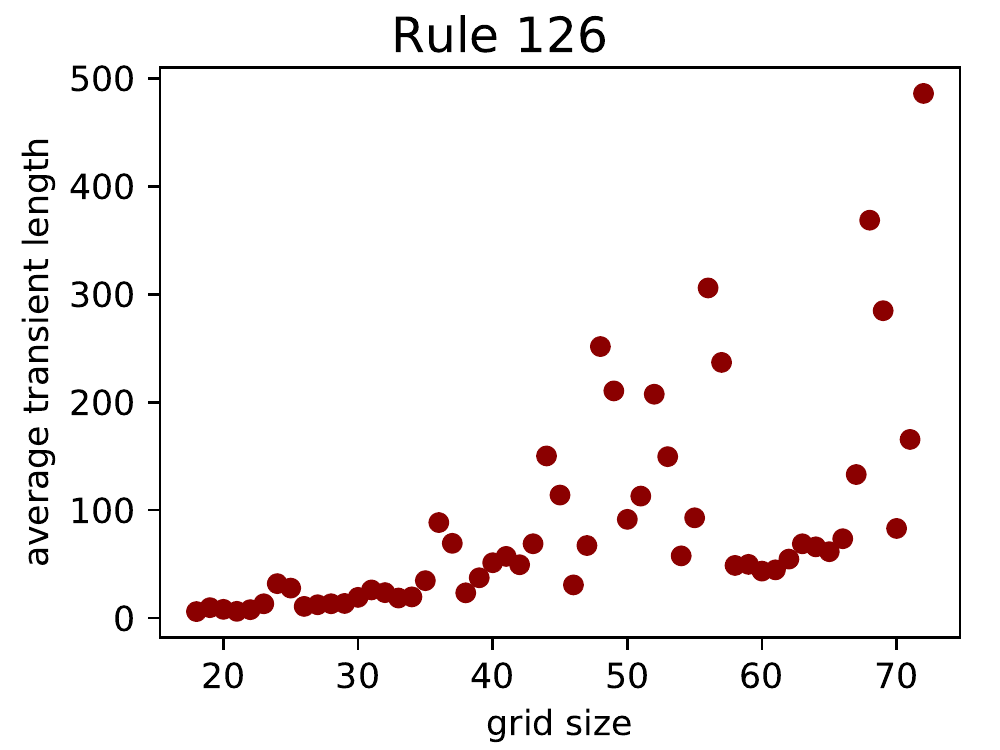}
  };

\end{tikzpicture}
\caption{Fractal Class rule 126. The average transient plot is on the left, the space-time diagram on the right.}
\end{figure}

\hyphenation{ap-proximated}

\subsection{Discussion}
We note that we have also tried to measure the asymptotic growth of the average attractor size $a_u$, $u \in \{0, 1 \}^n$ as well as the average rho value defined as $\rho_u = t_u + a_u$. This however produced data points, which could not be fitted to simple functions well. This is due to the fact that many automata have attractors consisting of a configuration, which is shifted by one bit to the left, resp. right, at every time step. The size of such an attractor then depends on the greatest common divisor of the size of the period of the attractor and the grid size, and this causes such oscillations. We conclude that such phase-space properties are not suitable for this classification method.

Below, we compare our results to other classifications described earlier. Exhaustive comparison for each ECA is presented in Table \ref{tab1}. 

 \tablefirsthead{\hline \multicolumn{5}{|c|}{\textbf{Classification Comparison}} \\ \hline
 \multicolumn{1}{|c||}{\textbf{ECA}} &  \multicolumn{1}{|c|}{\textbf{Transient}} &
  \multicolumn{1}{|c|}{\textbf{Wolfram}} & \multicolumn{1}{|c|}{\textbf{Zenil}} &  \multicolumn{1}{|c|}{\textbf{Wuensche}}
  \\ \hline}
   \tablehead{\hline \multicolumn{5}{|c|}{\textbf{Classification Comparison}} \\ \hline
 \multicolumn{1}{|c||}{\textbf{ECA}} &  \multicolumn{1}{|c|}{\textbf{Transient}} &
  \multicolumn{1}{|c|}{\textbf{Wolfram}} & \multicolumn{1}{|c|}{\textbf{Zenil}} &  \multicolumn{1}{|c|}{\textbf{Wuensche}}
  \\ \hline}
   \tabletail{\hline \multicolumn{5}{|r|}{{Continued in the subsequent column}} \\ \hline}
  \tablelasttail{\hline}
  \bottomcaption{Comparing classifications of the 88 unique ECAs.}
  \label{tab1}
\begin{center}
\begin{strictsupertabular}{@{\extracolsep{\fill}} |m{.8cm}||m{1.2cm}|m{.8cm}|m{1cm}|m{1cm}|  }  
0 & bounded & 1 &1 or 2  & 0\\
1 & bounded & 2 &1 or 2 & 0.25\\
2 & bounded & 2 & 1 or 2& 0.25\\
3 & bounded & 2 &1 or 2& 0.25\\
4 & bounded &2 &1 or 2& 0.25\\
5 & bounded & 2 & 1 or 2 & 0.5\\
6 & log & 2 & 1 or 2 & 0.5\\
7 & log & 2 &  1 or 2& 0.75\\
8 & bounded & 1&1 or 2& 0.25\\
9 & lin & 2 & 1 or 2 & 0.5\\
10 & bounded & 2 & 1 or 2 & 0.5\\
11 & log & 2 & 1 or 2 & 0.75\\
12 & bounded & 2 & 1 or 2 & 0.5\\
13 & log & 2 & 1 or 2 & 0.75\\
14 & lin & 2 &  1 or 2& 0.75\\
15 & bounded & 2 & 1 or 2 & 1\\
18 & fractal & 2/3 & 1 or 2 & 0.5\\
19 & bounded & 2 &  1 or 2& 0.625\\
22 & exp & 2/3 & 1 or 2 & 0.75\\
23 & log & 2 & 1 or 2 & 0.5\\
24 & bounded & 2 &  1 or 2& 0.5\\
25 & lin & 2 & 1 or 2 & 0.75\\
26 & log & 2 &  1 or 2& 0.75\\
27 & log & 2 & 1 or 2 & 0.75\\
28 & log & 2 &  1 or 2& 0.75\\
29 & bounded & 2 & 1 or 2 & 0.5\\
30 & exp & 3 &  3 & 1\\
32 & log & 1 & 1 or 2 & 0.25 \\
33 & log & 2 &  1 or 2& 0.5\\
34 & bounded & 2 & 1 or 2 & 0.5\\
35 & log & 2 &  1 or 2& 0.625\\
36 & bounded & 2 & 1 or 2 & 0.5\\
37 & log & 2 & 1 or 2 & 0.75\\
38 & bounded & 2 & 1 or 2 & 0.75\\
40 & log & 1 &  1 or 2& 0.5\\
41 & log & 2 & 1 or 2 & 0.75\\
42 & bounded & 2 & 1 or 2 & 0.75\\
43 & lin & 2 & 1 or 2 & 0.5\\
44 & log & 2 & 1 or 2 & 0.75\\
45 & exp & 3 &  3 & 1\\
46 & bounded & 2 & 1 or 2 & 0.5\\
50 & log & 2 & 1 or 2 & 0.625\\
51 & bounded & 2 &1 or 2 & 1\\
54 & exp & 2/4 &  1 or 2 & 0.75\\
56 & log & 2 & 1 or 2 & 0.75\\
57 & lin & 2 & 1 or 2 & 0.75\\
58 & log & 2 & 1 or 2 & 0.75\\
60 & affine & 2 & 1 or 2 & 1\\
62 & lin & 2 & 1 or 2 & 0.75\\
72 & bounded & 1 & 1 or 2 & 0.5\\
73 & exp & 3/4 & 3 & 0.75\\
74 & log & 2 & 1 or 2 & 0.75\\
76 & bounded & 2 &1 or 2 & 0.625\\
77 & log & 2 & 1 or 2& 0.5\\
78 & log & 2 & 1 or 2& 0.75\\
90 & affine & 2 &1 or 2 & 1\\
94 & log & 2 & 1 or 2 & 0.75\\
104 & log & 1 & 1 or 2& 0.75\\
105 & affine & 2 &1 or 2 & 1\\
106 & exp & 3  &1 or 2 & 1\\
108 & bounded &1&1 or 2 & 0.75\\
110 & lin & 4 & 4 & 0.75\\
122 & fractal & 2/3 & 1 or 2& 0.75\\
126 & fractal & 2/3 & 1 or 2& 0.5\\
128 & log & 1 & 1 or 2 & 0.25\\
130 & log & 2 & 1 or 2& 0.5\\
132 & log & 2&1 or 2& 0.5\\
134 & log & 2 &1 or 2& 0.75\\
136 & log & 1 &1 or 2& 0.5\\
138 & bounded & 2 &1 or 2 & 0.75\\
140 & log & 2 &1 or 2 & 0.625\\
142 & lin & 2 &1 or 2 & 0.5\\
146 & fractal & 2/3 &1 or 2 & 0.75\\
150 & affine & 2 &1 or 2 & 1\\
152 & log & 2 & 1 or 2 & 0.75\\
154 & bounded & 2/3  & 1 or 2& 1\\
156 & log & 2 &1 or 2 & 0.75\\
160 & log & 1 & 1 or 2 & 0.5\\
162 & log & 2 & 1 or 2& 0.75\\
164 & log & 2 & 1 or 2& 0.75\\
168 & log & 1 & 1 or 2 & 0.75\\
170 & bounded & 2 & 1 or 2& 1\\
172 & log & 2 & 1 or 2 & 0.75\\
178 & log & 2 & 1 or 2& 0.5\\
184 & lin & 2 & 1 or 2& 0.5\\
200 & bounded & 1 & 1 or 2 & 0.625\\
204 & bounded & 2 &1 or 2 & 1\\
232 & log & 1 & 1 or 2& 0.5\\
  \end{strictsupertabular}
\end{center}

\hyphenation{ap-proximated}

\paragraph{Wolfram's Classification - Discussion}
The significance of our results stems precisely from the fact that the Transient Classification corresponds to Wolfram's so well. As it is not clear for many rules, which Wolfram class they belong to, the main advantage is that we provide a formal criterion, upon which this could be decided.

In particular, rules in Classes Bounded and Log correspond to rules in either Class 1 or 2. Class Exp corresponds to the chaotic Class 3 and Class Lin contains Class 4 together with some Class 2 rules. We mention an interesting discrepancy: rule 54, which is possibly considered by Wolfram to be Turing complete, belongs to the Class Exp. This might suggest that computations performed by this rule can be on average quite inefficient.

\hyphenation{ap-proximated}
\paragraph{Zenil's Classification - Discussion}
Zenil's Classification of ECAs offers a great formalization of Wolfram's and seems to roughly correspond to it. Compared to the Transient Classification, it is however less fine grained. Moreover, it contains some arbitrary parameters such as the data representation and compression algorithm used. In addition, it uses a clustering technique, which requires data of multiple automata to be mutually compared in order to give rise to different classes. In contrast, the Transient Class can be determined for a single automaton without any context.

\hyphenation{ap-proximated}
\paragraph{Wuensche's Z-parameter - Discussion}
Wuensche suggests that complex behavior occurs around $Z = 0.75$, which agrees with the fact that Lin Class rules with steep slope (rule 110, 62, and 25) have precisely this $Z$ value. However, the $Z = 0.75$ is in fact quite frequent. This suggests that thanks to its simplicity, the $Z$ parameter can be used to narrow down a vast space of CA rules when searching for complexity. However, more refined methods have to be subsequently applied to find concrete CAs with interesting behavior.
\hyphenation{ap-proximated}

\section{Transient Classification of 2D CAs}
So far we have examined the toy model of ECAs. The true usefulness of the classification would stem from its application to more complex CAs where it could be used to discover automata with interesting behavior.

We therefore applied it on a subset of two-dimensional CAs with a $3 \times 3$ neighborhood and 3 states to see whether 2D automata would still exhibit such clear transient growths.

We consider the 2D CAs to operate on a finite square grid of size $n \times n$. We consider the topology of the grid to be that of a torus in order for each cell to have a uniform neighborhood. In such scenario, the definition of transients is analogous to the one-dimensional case.

To reduce the vast automaton space, we only considered such automata whose local rules are invariant to all the symmetries of a square. As there are still $3^{2861}$ such symmetrical 2D CAs, we randomly sampled 10 000 of them.

We estimated the average transient length analogously to the 1D case and measured the asymptotic growth with respect to $n$ - the size of the side of the square grid. This is motivated by the fact that in a $n \times n$ grid the greatest distance between two cells depends linearly on $n$ rather than quadratically. 

We were able to classify $93.03 \%$ of 10 000 sampled automata with a time bound of 40 seconds for the computation of one transient length value on a single CPU. We estimate that most CAs are unclassified due to such computation resources restriction or due to rather strict conditions we imposed on a good regression fit. We obtained the same major classes - the Log, Lin, and Exp Class. However, in this case, the Exp Class seems to dominate the rule space. Another interesting difference is that a new class was observed - the polynomial class - which contains rules whose transients grow approximately quadratically. Moreover, our results suggest that the occurence of bounded class CAs in 2D is much scarcer as we found no such CAs in our sample.

\begin{table}[h!]
\begin{tabular}{ |m{3cm}||m{4.6cm}|  }
 \hline
 \multicolumn{2}{|c|}{Classification of 2D 3-state CAs (10 000 samples)} \\
 \hline
 \centering
 Transient Class & \centering Percentage of CAs
 \tabularnewline
 \hline
Bounded Class & 0\%\\
Log Class &  18.21\%\\
Lin Class & 1.17\%\\
Poly Class & 1.03\%\\
Exp Class & 72.62\%\\
Unclassified & 6.97\%\\
 \hline
\end{tabular}
\caption{ Classification of 10 000 randomly sampled symmetric 2D 3-state CAs.}
\label{2D}
\end{table}

We observed the space-time diagrams of randomly sampled automata from each class to infer its typical behavior. On average, the Log Class automata quickly enter attractors of small size. Lin Class exhibit emergence of various local structures. For automata with more gradual incline, such structures seem to die out quite fast. However, automata with steeper slopes exhibit complex interactions of such structures. The Poly class automata with steep slope seem to produce spatially separated regions of chaotic behavior against a static background. In the case of more gradual slopes, some local structures emerge. Finally, the Exp Class seems to be evolving chaotically with no apparent local structures. We present various examples of CA evolution dynamics in the form of GIF animations here\footnotemark.
\footnotetext{\url{http://bit.ly/transient_classification}}

These observations suggest that the region of Lin Class with steep slope and Poly class with more gradual incline seems to contain a non-trivial ratio of automata with complex behavior. In this sense, the Transient Classification can assist us to automatically search for complex automata similarly to the method designed by \cite{hugo}, where interesting novel automata were discovered by measuring growth of structured complexity using a data compression approach.

\section{Transients Classification of Other Well Known CAs}
We were interested whether some well-known complex automata from larger CA spaces would conform to the transient classification as well. Surprisingly, the result is positive.

\begin{figure}[h!] 
\begin{tikzpicture}[thick, every node/.style={inner sep=0,outer sep=0}]
  \node at (4, +0.08) {
     \includegraphics[width=0.44\linewidth]{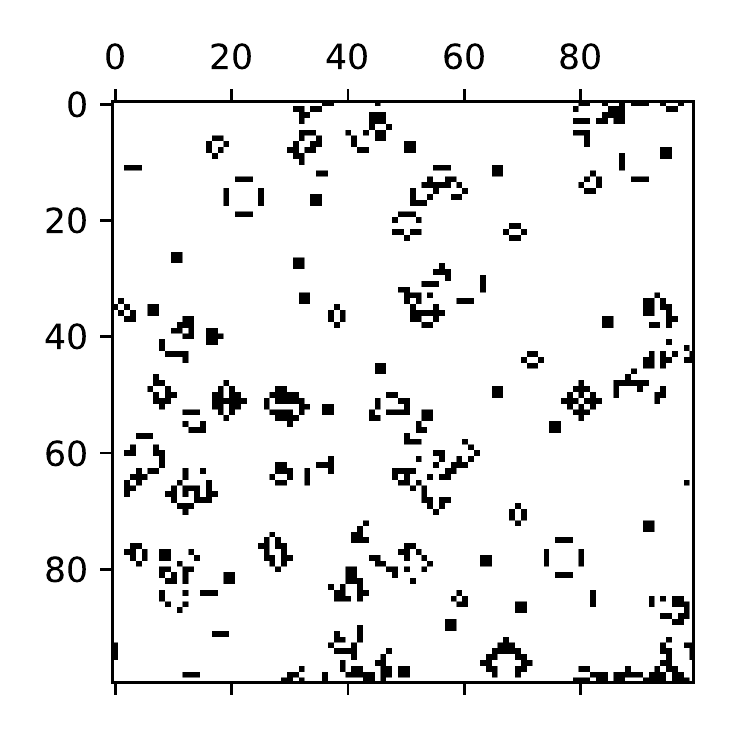}
  };
  \node at (-0.215, -0.1) {
     \includegraphics[width=0.6\linewidth]{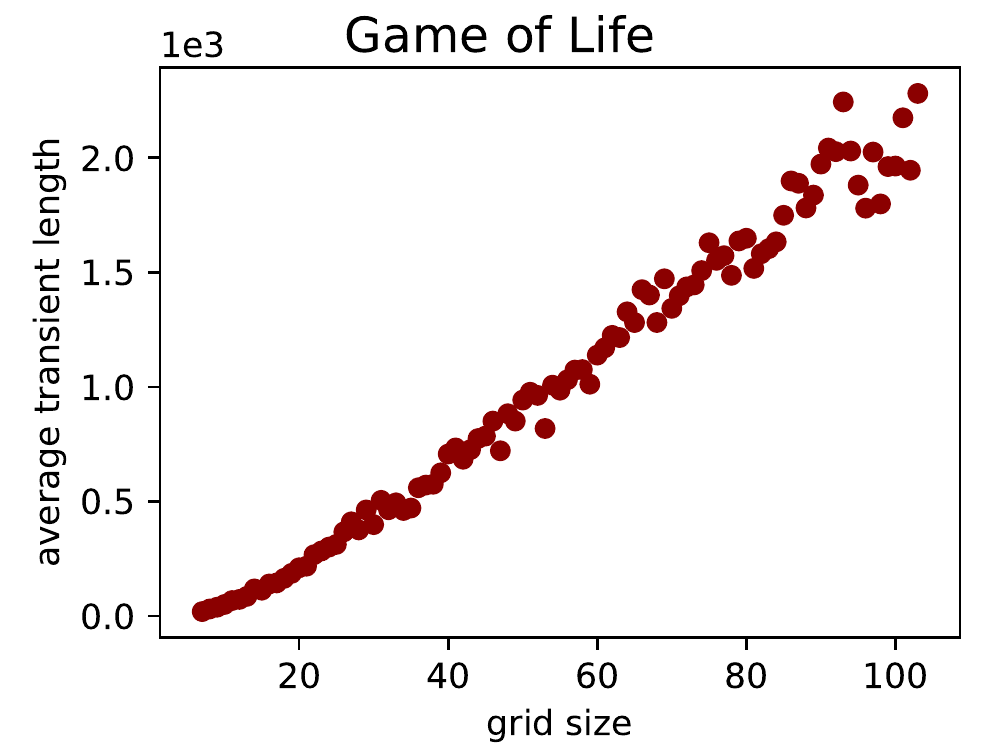}
  };
\end{tikzpicture}
\caption{Game of Life. The average transient growth plot is on the left. On the right, we show a space-time diagram at time $t=200$ started from a random initial configuration.}
\label{gol_pic}
\end{figure}

\paragraph{Game of Life} As the left plot in Figure \ref{gol_pic} suggests, the Turing complete Game of Life (\cite{game_of_life}) seems to fit the Lin Class. This is confirmed by the linear regression fit with $R^2 \approx 98.4\% $.

\paragraph{Genetically Evolved Majority CA} 
\cite{mitchellGA} studied how genetic algorithms can evolve CAs capable of global coordination. The authors were able to find a 1D CA denoted as $\phi_{par}$ with two states and radius $r=3$, which is quite successful at computing the majority task with the output required to be of the form of a homogenous state of either all 0's or all 1's. 

\begin{figure}[h!]
\begin{tikzpicture}[thick, every node/.style={inner sep=0,outer sep=0}]
  \node at (4, +0.08) {
     \includegraphics[width=0.44\linewidth]{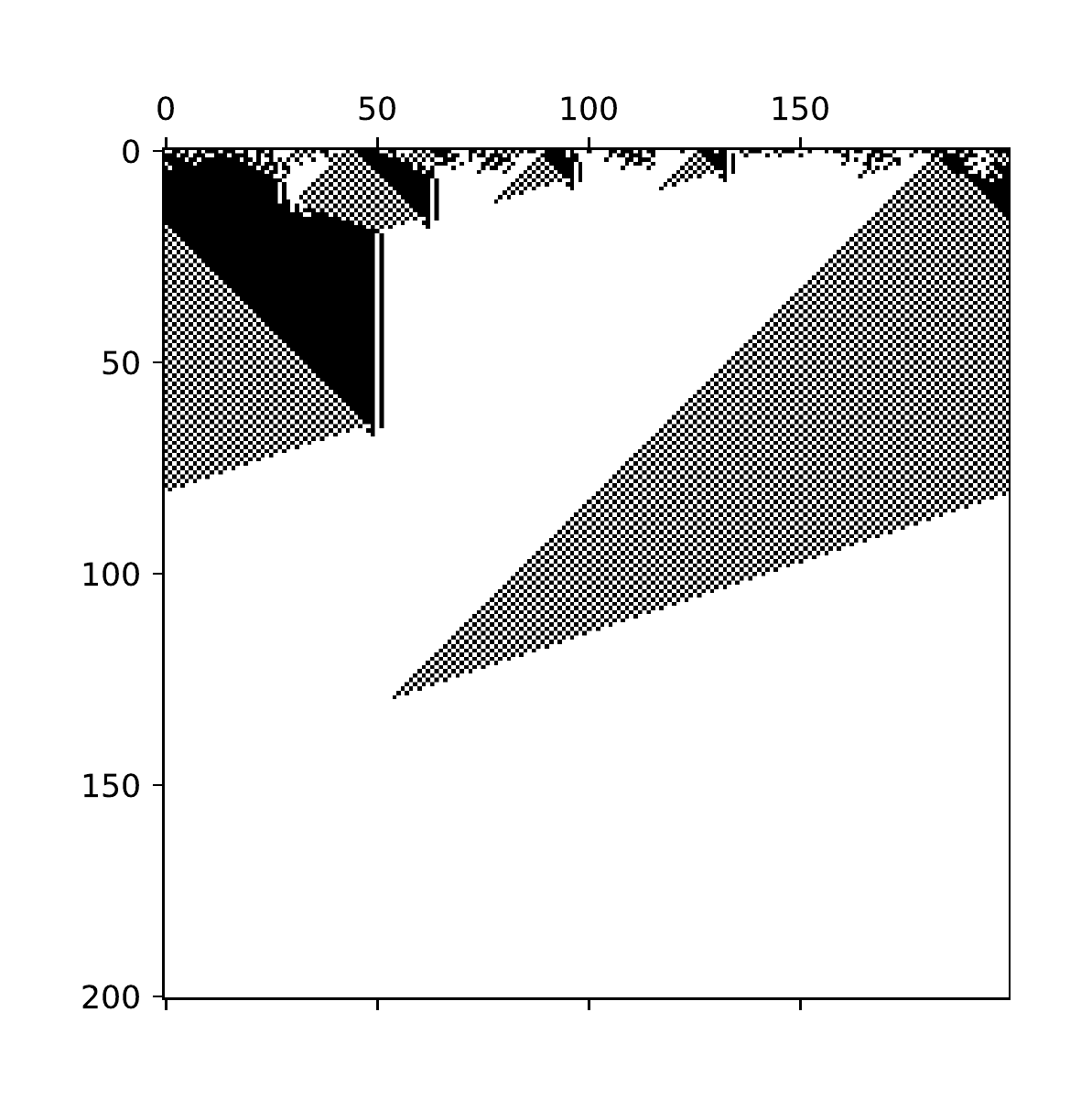}
  };
  \node at (-0.215, -0.1) {
     \includegraphics[width=0.6\linewidth]{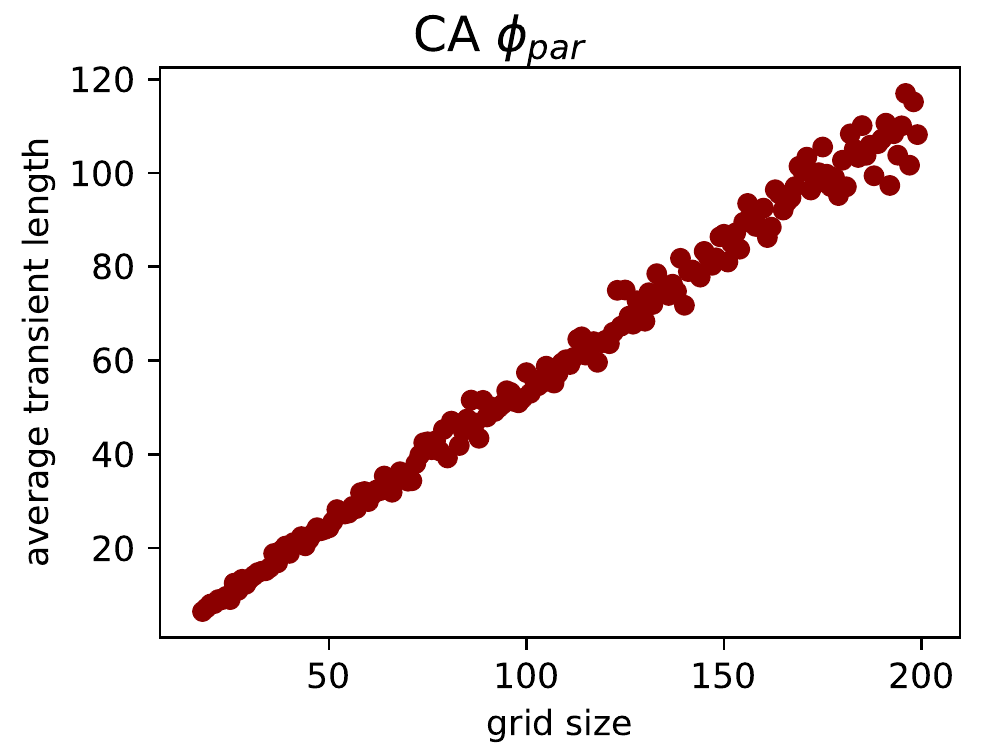}
  };

\end{tikzpicture}
\caption{Cellular automaton $\phi_{par}$. The average transient growth plot is on the left. On the right, we show a space-time diagram simulated from a random initial configuration.}
\end{figure}

This CA seems to belong to the Lin Class, which is confirmed by the linear regression fit with $R^2 \approx 99.2\% $.

\paragraph{Totalistic 1D 3-state CA} 
A totalistic CA is any CA whose local rule depends only on the number of cells in each state and not on their particular position. Wolfram studied various CA classes, one of them being the totalistic 1D CAs with radius $r=1$ and 3 states $S = \{0, 1, 2 \}$.

\begin{figure}[h!] 
\begin{tikzpicture}[thick, every node/.style={inner sep=0,outer sep=0}]
  \node at (4, +0.08) {
     \includegraphics[width=0.44\linewidth]{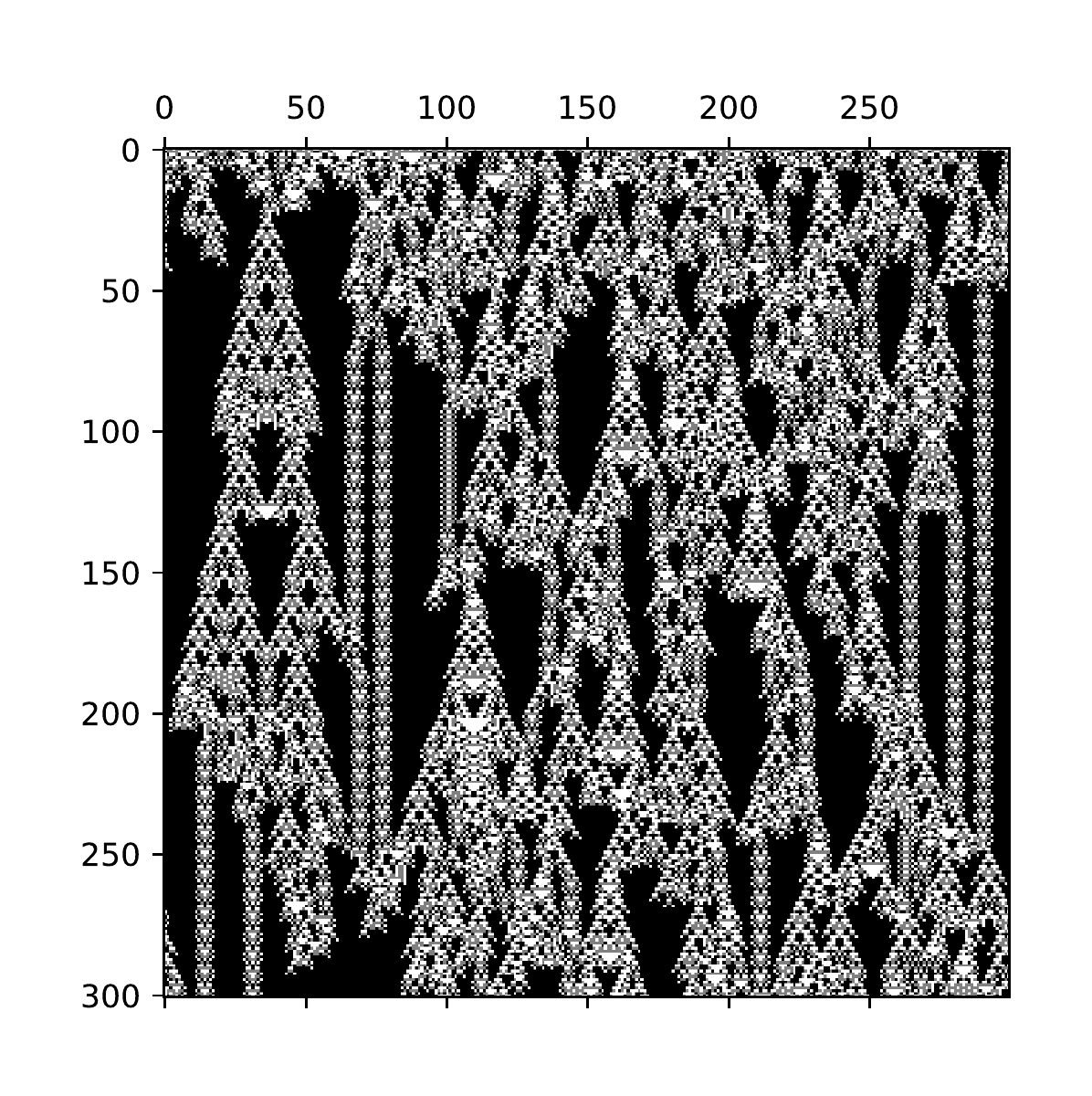}
  };
  \node at (-0.215, -0.1) {
     \includegraphics[width=0.6\linewidth]{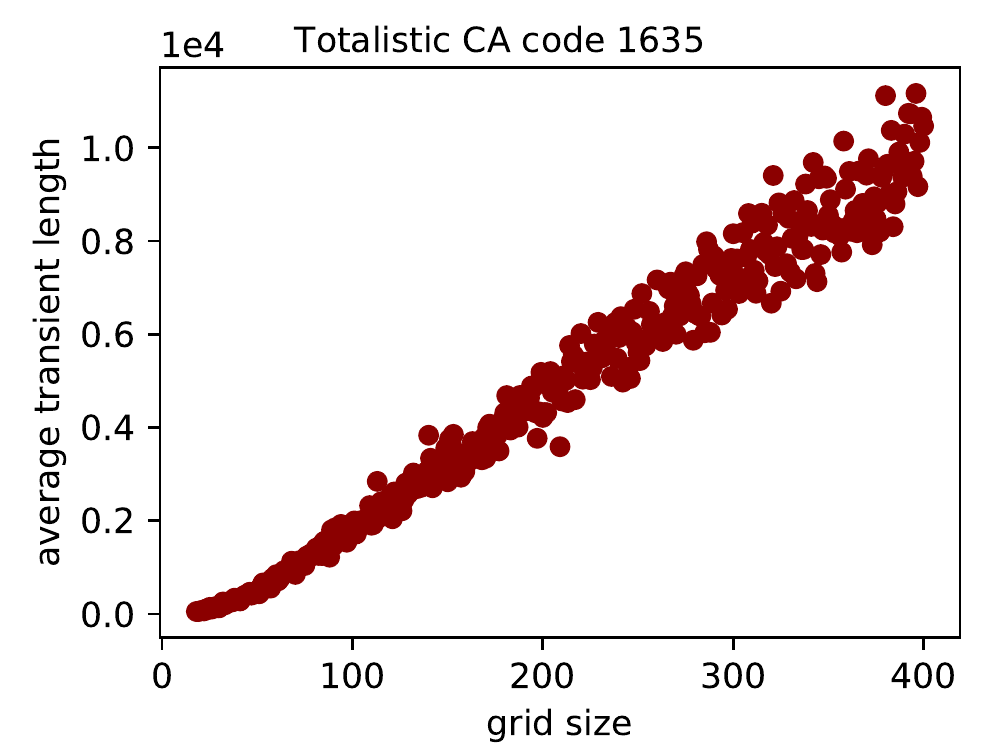}
  };

\end{tikzpicture}
\caption{Totalistic cellular automaton with code $1635$. The average transient growth plot is on the left. On the right, we show a space-time diagram of the evolution from a random initial configuration.}
\label{wolf_tot}
\end{figure}

\cite{newscience} presents a list of possibly complex CAs from this class. We applied the Transient Classification to such CAs and learned that most of them were classified as logarithmic. This agrees with our space-time diagram observations that the local structures in such CAs "die out" quite quickly. Nonetheless, some of the CAs were classified as linear. An example of such a CA is in Figure \ref{wolf_tot} where the linear regression fit has $R^2 \approx 97.63\% $.

\section{Conclusion}
We presented a classification method based on the asymptotic growth of average computation time, which is applicable to any deterministic discrete space and time dynamical system. We did present a good correspondence between the Transient and Wolfram's classification in the case of ECAs. We also did show that the transient classification works in two dimensions and used it to discover CAs capable of emergent phenomena. By demonstrating that famous CAs such as Game of Life or rule 110 belong to the Lin Class, we believe that the linear transient growth navigates us toward a region of complex and interesting CAs.

Our future work includes publishing an open-source library with the classification techniques. We also plan to compare the classification results for 2D CAs with various neighborhoods and number of states to compare, which ones have the largest ratio of the Lin and Poly Class automata. We would also like to discover, which other types of discrete dynamical systems can this method be applied to.

\section{Acknowledgements}
We would like to thank Jiri Tuma for all his help and support as well as Jaromir Antoch, Ondrej Tybl, and Hugo Cisneros for numerous inspiring discussions.

\footnotesize
\bibliographystyle{apalike}
\bibliography{example} 

\begin{thebibliography}{}

\bibitem[Bennett, 1988]{bennett}
Bennett, C.~H. (1988).
\newblock Logical depth and physical complexity.
\newblock {\em The Universal Turing Machine -- a Half-Century Survey}, pages
  227--257.

\bibitem[Chaitin, 1966]{chaitin}
Chaitin, G.~J. (1966).
\newblock On the length of programs for computing finite binary sequences.
\newblock {\em J. ACM}, 13(4):547–569.

\bibitem[Cisneros et~al., 2019]{hugo}
Cisneros, H., Sivic, J., and Mikolov, T. (2019).
\newblock Evolving structures in complex systems.
\newblock {\em Proceedings of the 2019 IEEE Symposium Series on Computational
  Intelligence}, pages 230--238.

\bibitem[Cook, 2004]{cook}
Cook, M. (2004).
\newblock Universality in elementary cellular automata.
\newblock {\em Complex Systems}, 15.

\bibitem[Crutchfield and Young, 1989]{statistical_complexity}
Crutchfield, J. and Young, K. (1989).
\newblock Inferring statistical complexity.
\newblock {\em Physical review letters}, 63:105--108.

\bibitem[Crutchfield and Hanson, 1993]{crutchfield}
Crutchfield, J.~P. and Hanson, J.~E. (1993).
\newblock Turbulent pattern bases for cellular automata.
\newblock {\em Physica D: Nonlinear Phenomena}, 69(3):279 -- 301.

\bibitem[Gardener, 1970]{game_of_life}
Gardener, M. (1970).
\newblock The fantastic combinations of {J}ohn {C}onway’s new solitaire game
  “life” by {M}artin {G}ardner.
\newblock {\em Scientific American}, 223:120--123.

\bibitem[Gutowitz et~al., 1987]{gutowitz_local}
Gutowitz, H.~A., Victor, J.~D., and Knight, B.~W. (1987).
\newblock Local structure theory for cellular automata.
\newblock {\em Physica D: Nonlinear Phenomena}, 28(1):18 -- 48.

\bibitem[Hanson, 2009]{emergence1}
Hanson, J.~E. (2009).
\newblock {Emergent Phenomena in Cellular Automata}.
\newblock {\em {Meyers R. (eds) Encyclopedia of Complexity and Systems
  Science}}.

\bibitem[Hedlund, 1969]{hedlund}
Hedlund, G.~A. (1969).
\newblock Endomorphisms and automorphisms of the shift dynamical system.
\newblock {\em Mathematical systems theory}, 3:320--375.

\bibitem[Kaneko, 1985]{kaneko}
Kaneko, K. (1985).
\newblock Complexity in basin structures and information processing by the
  transition among attractors.
\newblock {\em Theory and Applications of Cellular Automata}, pages 367--399.

\bibitem[Kari, 2005]{Kari_survey}
Kari, J. (2005).
\newblock Theory of cellular automata: A survey.
\newblock {\em Theoretical Computer Science}, 334(1):3 -- 33.

\bibitem[Langton, 1984]{LANGTONselfrep}
Langton, C.~G. (1984).
\newblock Self-reproduction in cellular automata.
\newblock {\em Physica D: Nonlinear Phenomena}, 10(1):135 -- 144.

\bibitem[Martin et~al., 1984]{algCAs}
Martin, O., Odlyzko, A., and Wolfram, S. (1984).
\newblock Algebraic properties of cellular automata.
\newblock {\em Communications in Mathematical Physics}, 93.

\bibitem[McShea, 1996]{mcshea}
McShea, D. (1996).
\newblock Perspective: Metazoan complexity and evolution: Is there a trend?
\newblock {\em Evolution}, 50.

\bibitem[Mitchell, 1998]{mitchell_overview}
Mitchell, M. (1998).
\newblock Computation in cellular automata: A selected review.
\newblock {\em Non‐Standard Computation}, pages 95--140.

\bibitem[Mitchell et~al., 2000]{mitchellGA}
Mitchell, M., Crutchfield, J., and Das, R. (2000).
\newblock Evolving cellular automata with genetic algorithms: A review of
  recent work.
\newblock {\em First Int. Conf. on Evolutionary Computation and Its
  Applications}, 1.

\bibitem[Neumann and Burks, 1966]{neuman}
Neumann, J.~V. and Burks, A.~W. (1966).
\newblock {\em Theory of Self-Reproducing Automata}.
\newblock University of Illinois Press, Urbana, USA.

\bibitem[Ofria and Wilke, 2004]{avida}
Ofria, C. and Wilke, C. (2004).
\newblock Avida: A software platform for research in computational evolutionary
  biology.
\newblock {\em Artificial Life}, 10(2):191--229.

\bibitem[Owen, 2013]{mcbook}
Owen, A.~B. (2013).
\newblock {\em Monte Carlo theory, methods and examples}.

\bibitem[Ray, 1991]{tierra}
Ray, T.~S. (1991).
\newblock An approach to the synthesis of life.
\newblock {\em Artificial Life II, Santa Fe Institute Studies in the Sciences
  of Complexity}, XI:371–408.

\bibitem[Reggia et~al., 1993]{REGGIAselfrep}
Reggia, J., Armentrout, S., Chou, H., and Peng, Y. (1993).
\newblock Simple systems that exhibit self-directed replication.
\newblock {\em Science (New York, N.Y.)}, 259:1282--7.

\bibitem[Saclay and Gutowitz, 1994]{gutowitz}
Saclay, C. and Gutowitz, H. (1994).
\newblock Transients, cycles, and complexity in cellular automata.
\newblock {\em Physical Review A}, 44.

\bibitem[Soros and Stanley, 2014]{chromaria}
Soros, L. and Stanley, K. (2014).
\newblock Identifying necessary conditions for open-ended evolution through the
  artificial life world of chromaria.
\newblock {\em Artificial Life Conference Proceedings}, (26):793--800.

\bibitem[Stepney, 2012]{natural_comp}
Stepney, S. (2012).
\newblock {\em Nonclassical Computation --- A Dynamical Systems Perspective}.
\newblock Springer Berlin Heidelberg, Berlin, Heidelberg.

\bibitem[Toffoli, 1977]{toffoli}
Toffoli, T. (1977).
\newblock Computation and construction universality of reversible cellular
  automata.
\newblock {\em Journal of Computer and System Sciences}, 15(2):213 -- 231.

\bibitem[Vichniac, 1984]{vichniac}
Vichniac, G.~Y. (1984).
\newblock Simulating physics with cellular automata.
\newblock {\em Physica D: Nonlinear Phenomena}, 10(1):96 -- 116.

\bibitem[Wolfram, 2002]{newscience}
Wolfram, S. (2002).
\newblock {\em A New Kind of Science}.
\newblock Wolfram Media, Champaign, USA.

\bibitem[Wuensche, 2016]{ddlab}
Wuensche, A. (2016).
\newblock {\em Exploring discrete dynamics - Second Edition. The DDLab manual}.
\newblock Luniver Press.

\bibitem[Wuensche and Lesser, 2001]{wuensche_global}
Wuensche, A. and Lesser, M. (2001).
\newblock The global dynamics of celullar automata: An atlas of basin of
  attraction fields of one-dimensional cellular automata.
\newblock {\em J. Artificial Societies and Social Simulation}, 4.

\bibitem[Zenil, 2009]{zenil_compression}
Zenil, H. (2009).
\newblock Compression-based investigation of the dynamical properties of
  cellular automata and other systems.
\newblock {\em Computing Research Repository - CORR}, 19.

\end{thebibliography}

\end{document}